\documentclass[
 aps,
 prx,
 amsmath,amssymb,
 reprint,
 10pt,
 superscriptaddress,
 floatfix,
 nofootinbib
]{revtex4-2}

\usepackage{physics}

\usepackage[english]{babel}
\usepackage[utf8]{inputenc}
\usepackage[T1]{fontenc}
\usepackage{mathrsfs}
\usepackage{bbm}
\usepackage{bm}
\usepackage{enumerate}
\usepackage{graphicx}
\usepackage[dvipsnames]{xcolor}
\usepackage{float}
\usepackage{subdepth}
\usepackage{fnpct}%multiple footnotes
\usepackage{booktabs}
\usepackage{multirow}
\usepackage[math]{cellspace}

\usepackage{lipsum}
\usepackage{comment}

\usepackage[colorlinks,
linkcolor=BrickRed,
citecolor=MidnightBlue,
urlcolor=MidnightBlue,
bookmarks=true,        
bookmarksopen=true,    
bookmarksnumbered=true,
]{hyperref}

\renewcommand*\d{\mathop{}\!\mathrm{d}}

\newcommand{\pd}{\partial}

\renewcommand{\i}{\mathrm{i}}
\renewcommand{\(}{\left(}
\renewcommand{\)}{\right)}
\newcommand{\id}{\mathbbm{1}}

\newcommand{\scT}{\mathcal{T}}
\newcommand{\scC}{\mathcal{C}}
\newcommand{\scP}{\mathcal{P}}
\newcommand{\scQ}{\mathcal{Q}}
\newcommand{\scU}{\mathcal{U}}

\newcommand{\scI}{\mathcal{I}}
\newcommand{\scK}{\mathcal{K}}
\newcommand{\scL}{\mathcal{L}}
\newcommand{\scA}{\mathcal{A}}
\newcommand{\scLH}{\mathcal{L}_{\mathrm{H}}}
\newcommand{\scLD}{\mathcal{L}_{\mathrm{D}}}
\newcommand{\scLJ}{\mathcal{L}_{\mathrm{J}}}
\newcommand{\swap}{\mathcal{S}}
\newcommand{\tphi}{\tilde{\phi}}
\newcommand{\balpha}{{\overline{\alpha}}}

\newcommand{\matAIp}{
	\begin{pmatrix} & A \\ B & \end{pmatrix},\ \begin{cases}A=A^*\\B=B^*\end{cases}
}
\newcommand{\matAIm}{
	\begin{pmatrix} & A \\ A^* & \end{pmatrix}
}

\newcommand{\matBDId}{
	\begin{pmatrix}
		A & B \\
		B^\top &  C
	\end{pmatrix},\
	\begin{cases}
		A=A^\dagger=A^*=A^\top\\
		B=-B^*\\
		C=C^\dagger=C^*=C^\top\\
	\end{cases}
}

\newcommand{\matBDIalt}{
	\begin{pmatrix}
		A & B \\
		-B^\top &  C
	\end{pmatrix},\
	\begin{cases}
		A=-A^\dagger=-A^\top=A^*\\
		B=-B^*\\
		C=-C^\dagger=-C^\top=C^*
	\end{cases}
}

\newcommand{\matCIalt}{
	\begin{pmatrix}
		A & B \\
		-B^* &  -A^*
	\end{pmatrix},\
	\begin{cases}
		A=A^\dagger\\
		B=B^\top
	\end{cases}
}

\newcommand{\matBDIdpp}{
	\begin{pmatrix}
		& & A & B \\
		& & C &  D \\
		A^\top & C^\top & & \\
		B^\top & D^\top & &
	\end{pmatrix},\
	\begin{cases}
		A=A^*\\
		B=-B^*\\
		C=-C^*\\
		D=D^*
	\end{cases}
}

\newcommand{\matBDIdmm}{
	\begin{pmatrix}
		& & A & B \\
		& & B^\top &  C \\
		A^* & -B^* & & \\
		-B^\dagger & C^* & &
	\end{pmatrix},\
	\begin{cases}
		A=A^\top\\
		C=C^\top
	\end{cases}
}

\newcommand{\matBDIdpm}{
	\begin{pmatrix}
		 & A \\
		A^\top &  
	\end{pmatrix},\
	A=A^\dagger
}

\newcommand{\matBDIdmp}{
	\begin{pmatrix}
		& A \\
		B &  
	\end{pmatrix},\
	\begin{cases}
		A=A^\dagger=A^\top=A^*\\
		B=B^\dagger=B^\top=B^*
	\end{cases}
}

\newcounter{ls}

\newcounter{pr}

\newcounter{tpcom}

\begin{document}
	
\title{Symmetry Classification of Many-Body Lindbladians: Tenfold Way and Beyond}

\author{Lucas S\'a}
\email{lucas.seara.sa@tecnico.ulisboa.pt}

\affiliation{CeFEMA, Instituto Superior T\'ecnico, Universidade de Lisboa, Av.\ Rovisco Pais, 1049-001 Lisboa, Portugal}

\author{Pedro Ribeiro}
\email{ribeiro.pedro@tecnico.ulisboa.pt}

\affiliation{CeFEMA, Instituto Superior T\'ecnico, Universidade de Lisboa, Av.\ Rovisco Pais, 1049-001 Lisboa, Portugal}
\affiliation{Beijing Computational Science Research Center, Beijing 100193, China}

\author{Toma\v z Prosen}
\email{tomaz.prosen@fmf.uni-lj.si}
\affiliation{Department of Physics, Faculty of Mathematics and Physics, University of Ljubljana, SI-1000 Ljubljana, Slovenia}

\begin{abstract}
We perform a systematic symmetry classification of many-body Lindblad superoperators describing general (interacting) open quantum systems coupled to a Markovian environment. Our classification is based on the behavior of the many-body Lindbladian under antiunitary symmetries and unitary involutions. We find that Hermiticity preservation reduces the number of symmetry classes, while trace preservation and complete positivity do not, and that the set of admissible classes depends on the presence of additional unitary symmetries: in their absence or in symmetry sectors containing steady states, many-body Lindbladians belong to one of ten non-Hermitian symmetry classes; if however, there are additional symmetries and we consider non-steady-state sectors, they belong to a different set of 19 classes. In both cases, it does not include classes with Kramers degeneracy. Remarkably, our classification admits a straightforward generalization to the case of non-Markovian, and even non-trace-preserving, open quantum dynamics. While the abstract classification is completely general, we then apply it to general (long-range, interacting, spatially inhomogeneous) spin-$1/2$ chains. We explicitly build examples in all ten classes of Lindbladians in steady-state sectors, describing standard physical processes such as dephasing, spin injection and absorption, and incoherent hopping, thus illustrating the relevance of our classification for practical physics applications. Finally, we show that the examples in each class display unique random-matrix correlations. To fully resolve all symmetries, we employ the combined analysis of bulk complex spacing ratios and the overlap of eigenvector pairs related by symmetry operations. We further find that statistics of levels constrained onto the real and imaginary axes or close to the origin are not universal due to spontaneous breaking of Lindbladian PT symmetry.
\end{abstract}

\maketitle

\section{Introduction}
The interplay of symmetries, correlations, and dynamics lies at the heart of our understanding of complex interacting quantum many-body systems. It provides a compact and powerful framework for obtaining universal information not otherwise available for generic quantum systems.
Hamiltonians are classified by a reduced number of global antiunitary symmetries and unitary involutions. The behavior under time-reversal, particle-hole, and chiral symmetry places them in one of the ten celebrated Altland-Zirnbauer classes~\cite{altland1997}. In turn, the Bohigas-Giannoni-Schmit conjecture~\cite{bohigas1984} states that, if the system is chaotic, the Hamiltonian displays the statistical behavior of a random matrix from the same symmetry class. Finally, the correlations of random matrices are universal and solely determined by its symmetry class: level repulsion is a direct measure of the system's behavior under time reversal, while the spectral density close to the origin is determined by particle-hole and chiral symmetry~\cite{verbaarschot1993,verbaarschot1994,akemann1997}. Quantities such as conductance fluctuations in disordered electronic systems can thus be inferred solely from the knowledge of invariance under simple symmetry transformations~\cite{altland1997}.

Recent years have seen a revival of interest in non-Hermitian physics~\cite{ashida2020}, which is of relevance, for instance, in PT-symmetric, dissipative, and monitored quantum dynamics, and also in classical and optical setups. However, the study of symmetries and correlations of non-Hermitian quantum matter is much less developed than its Hermitian counterpart. Non-Hermiticity bifurcates time-reversal and particle-hole symmetries into two distinct types each, while pseudo- and anti-pseudo-Hermiticity are additional transformations of the Hamiltonian. The symmetry classification of non-Hermitian Hamiltonians was only recently settled~\cite{kawabata2019,zhou2019}, when it was found that there are 38 non-Hermitian symmetry classes, the so-called Bernard-LeClair classes~\cite{bernard2002}, for point-gap spectra (i.e., spectra that can be freely rotated in the complex plane) and 54 classes~\cite{liu2019PRB,ashida2020} for line-gap spectra (which cannot). Similarly to the Hermitian case, there are three classes of universal bulk level repulsion~\cite{hamazaki2019}, determined by the behavior under transposition. Long-range correlations can be understood in terms of the dissipative spectral form factor~\cite{fyodorov1997,li2021PRL,shivam2022} and its local deformations~\cite{garcia2022ARXIV}. The statistics on, or near, the real axis for real asymmetric matrices are also well understood~\cite{lehmann1991PRL,kanzieper2005PRL,forrester2007PRL}. In general, analytical results are only available for the three standard and the three chiral Ginibre classes~\cite{ginibre1965,halasz1997,osborn2004PRL,akemann2009PRE,akemann2009JPhysA,akemann2011Acta,dusa2022}, although there is an increasing body of numerical results for the many other classes~\cite{hamazaki2019PRL,kanazawa2021,garcia2022PRX,xiao2022,hamazaki2022,gosh2022}.

Perhaps more importantly, non-Hermitian Hamiltonians provide an effective description of open quantum dynamics only when quantum jumps can be neglected, for instance, for short times or postselecting jump-free quantum trajectories. A complete description of an open quantum system coupled to a Markovian (i.e., memoryless) environment must go beyond the non-Hermitian Hamiltonian description, and one should consider systems evolving under the action of Liouvillian superoperators of Lindblad form~\cite{belavin1969,gorini1976,lindblad1976} (Lindbladians for short). It is a question of fundamental interest to find out how many symmetry classes can be realized by many-body Lindbladians, which are far more constrained than arbitrary non-Hermitian Hamiltonians, specifically by the conservation of trace, Hermiticity, and (complete) positivity. In other words, we ask to which subset of the 54-fold classification do physical open quantum systems belong. Lieu, McGinley, and Cooper~\cite{lieu2020} used causality arguments to argue that there are also ten classes of single-particle spectra of noninteracting (quadratic) Lindbladians. However, they did not consider shifting the spectral origin, which avoids the causality restrictions, as pointed out by Kawasaki, Mochizuki, and Obuse~\cite{kawasaki2022}. Once this possibility is accounted for, all 54 classes of non-Hermitian Hamiltonians can be implemented at the level of single-particle spectra. The importance of the shift of the spectral origin, and the associated spectral dihedral symmetry, was already noted for many-body Lindbladians by one of us in Refs.~\cite{prosen2012PRL,prosen2012PRA}, but a symmetry classification was not put forward. 

In this paper, we take this fundamental step and show that many-body Lindbladians possess a rich symmetry classification: in the absence of unitary symmetries or in symmetry sectors containing the steady state(s), they belong to one of {\em ten} non-Hermitian symmetry classes; if however, there are additional unitary symmetries and we consider non-steady-state sectors, they belong to a different set of 19 classes. Remarkably, our classification does not include any classes with Kramers degeneracy. It is remarkable to observe that the number of distinct symmetry classes of Lindbladian dynamics
in symmetry sectors that contain the steady state(s) is exactly the same (ten) as the number of distinct Altland-Zirnbauer symmetry classes of Hermitian steady-state density operators, although the precise correspondence remains to be understood.

Our work is qualitatively different from the previous attempt at a symmetry classification of fermionic open quantum matter by Altland, Diehl, and Fleischhauer~\cite{altland2021}, on the level of both generality and abstraction. 
Specifically, Ref.~\cite{altland2021} considers the invariance of the dynamics under linear or antilinear and canonical or anticanonical transformations of fermionic creation and annihilation operators, while our transformations apply to any kind of Hilbert space (including second-quantized Lindbladians in Fock space) and are defined by general transformation properties of the \emph{matrix representation} of the Lindbladian. As such, our classification scheme accurately captures many-body spectral and eigenvector properties, as relevant, e.g., for quantum chaos.

The goal of this paper is threefold. First, we establish the symmetry classification of many-body Lindbladians and determine conditions that the Hamiltonian and jump operators must satisfy in a given class (Sec.~\ref{sec:classification}).
Second, we propose several experimentally realizable examples of physical Lindbladians belonging to the full Lindbladian tenfold way (Sec.~\ref{sec:examples}), illustrating the practical relevance of our abstract classification. 
Third, we advance the understanding of non-Ginibre random-matrix ensembles, by proposing the eigenvector overlap matrix~\cite{mehlig1998PRL} as a detector of antiunitary symmetries and studying its statistical properties (Sec.~\ref{sec:rmt}). Specifically, we demonstrate that the
overlap matrix element between symmetry-related eigenstates together with the complex spacing ratio~\cite{sa2019csr} provides a unique indicator for the classification.

\section{Lindbladian symmetry classification}
\label{sec:classification}

\subsection{Matrix representation of the Lindbladian}

We consider the quantum master equation for the system's density matrix, $\pd_t\rho=\scL \rho$, where the Liouvillian superoperator is of the Lindblad form,
\begin{equation}
\scL \rho= -\i \comm{H}{\rho}+\sum_{m=1}^{M}\(2L_m\rho L_m^\dagger-\acomm{L_m^\dagger L_m}{\rho}\),
\label{lindf}
\end{equation}
with Hamiltonian $H$ and $M$ traceless jump operators $L_m$, $m=1,\dots,M$ acting over a Hilbert space $\mathcal{H}$. The Lindbladian admits a matrix representation (vectorization) over a doubled Hilbert space $\mathcal{H}\otimes\mathcal{H}$ (the so-called Liouville space), $\scL=\scLH+\scLD+\scLJ$, where the Hamiltonian, dissipative, and jump contributions are, respectively, given by
\begin{align}
\label{eq:scLH}
\scLH &= -\i \(H\otimes \id -\id \otimes H^*\),
\\
\label{eq:scLD}
\scLD&=-\(
\sum_{m=1}^M L_m^\dagger L_m \otimes \id 
+ \id \otimes \sum_{m=1}^M \(L_m^\dagger L_m\)^*
\),
\\
\label{eq:scLJ}
\scLJ&=2\sum_{m=1}^M  L_m\otimes L_m^*.
\end{align}
$()^*$ denotes complex conjugation in a matrix representation with respect to a fixed basis of $\mathcal H$ (or 
$\mathcal H\otimes \mathcal H$).
We see below that the three contributions have different transformation properties. We further define the traceless shifted Lindbladian~\cite{prosen2012PRL},
\begin{equation}\label{eq:scL_prime}
\scL'=\scL-\alpha\,\mathcal{I},
\qquad
\alpha=\frac{\Tr \scL}{\Tr \scI}=
-2\,\frac{\sum_m \Tr L_m^\dagger L_m}{\Tr \id},
\end{equation} 
where $\scI=\id\otimes \id$ is the identity operator over the Liouville space. As we show below, the symmetry classification of Lindbladians is necessarily formulated in terms of $\scL'$.

\subsection{Superoperator symmetries}

Just as for the Hamiltonian case, the symmetry classification of the Lindbladian follows from the behavior of its irreducible blocks under involutive antiunitary (superoperator) symmetries. More precisely, if there is a unitary superoperator $\scU$ that commutes with the Lindbladian $\scL$,
\begin{equation}
\scU\scL\scU^{-1}=\scL,
\end{equation}
we can block diagonalize (reduce) $\scL$ into sectors of fixed eigenvalues of $\scU$. For the moment, let us assume no such unitary symmetries exist and the Lindbladian is irreducible; we consider unitary symmetries in Sec.~\ref{subsec:unitary_syms}. We look for the existence of antiunitary superoperators $\scT_\pm$, such that $\scL$ satisfies
\begin{alignat}{99}
\label{eq:nHsym_Tp}
&\scT_+ \scL \scT_+^{-1} = +\scL,\qquad 
&&\scT_+^2=\pm 1,
\\
\label{eq:nHsym_Tm}
&\scT_- \scL \scT_-^{-1} = -\scL,\qquad 
&&\scT_-^2=\pm 1.
\end{alignat}
Since $\scL$ is non-Hermitian, it can also be related to its adjoint through antiunitary superoperators. To this end, we look for the existence of antiunitaries $\scC_\pm$ implementing:
\begin{alignat}{99}
\label{eq:nHsym_Cp}
&\scC_+ \scL^\dagger \scC_+^{-1} = +\scL,\qquad 
&&\scC_+^2=\pm 1,
\\
\label{eq:nHsym_Cm}
&\scC_- \scL^\dagger\scC_-^{-1} = -\scL,\qquad 
&&\scC_-^2=\pm 1.
\end{alignat}
We need not consider the existence of more than one antiunitary of a given kind, since their product is unitary and commutes with $\scL$ while we assumed $\scL$ to be irreducible. On the other hand, the combined action of antiunitaries of different types gives rise to new unitary involutions. 
In the absence of antiunitary symmetries, these unitary involutions can still act on their own and we look for unitary superoperators $\scP$ and $\scQ_\pm$, such that $\scL$ transforms as
\begin{alignat}{99}
\label{eq:nHsym_P}
&\scP \scL \scP^{-1} = -\scL,\qquad 
&&\scP^2=1,
\\
\label{eq:nHsym_Qp}
&\scQ_+ \scL^\dagger \scQ_+^{-1} =  +\scL,\qquad 
&&\scQ_+^2=1,
\\
\label{eq:nHsym_Qm}
&\scQ_- \scL^\dagger \scQ_-^{-1} = - \scL,\qquad 
&&\scQ_-^2=1.
\end{alignat}
Furthermore, the unitary involutions can either commute or anticommute with each other and with the antiunitary symmetries; that is,
\begin{align}
\label{eq:eps_1}
&\scP \scT =\epsilon_{\scP\!\scT}\ \scT \scP,
\quad
&\scP \scC =\epsilon_{\scP\scC}\ \scC \scP,
\\
&\scQ \scT =\epsilon_{\scQ\!\scT}\ \scT \scQ,
\quad
&\scQ \scC =\epsilon_{\scQ\scC}\ \scC \scQ,
\\
\label{eq:eps_5}
&\scQ \scP =\epsilon_{\scP\!\scQ}\ \scP \scQ,
\end{align}
where all $\epsilon=\pm1$ and $\scT$, $\scC$, and $\scQ$ can be one of $\scT_\pm$, $\scC_\pm$, or $\scQ_\pm$, respectively. Only three of the $\epsilon$ are independent, say, $\epsilon_{\scP\!\scT}$, $\epsilon_{\scQ\!\scT}$, and $\epsilon_{\scP\!\scQ}$. The remaining two are determined by $\epsilon_{\scP\scC}=\epsilon_{\scP\!\scQ}\epsilon_{\scP\!\scT}$ and $\epsilon_{\scQ\scC}=\epsilon_{\scQ\!\scT}$.

The symmetries of Eqs.~(\ref{eq:nHsym_Tp})--(\ref{eq:nHsym_Qm}) describe two independent flavors of time-reversal ($\scT_+$ and $\scC_+$) and particle-hole ($\scT_-$ and $\scC_-$) symmetries, chiral or sublattice symmetry ($\scP$) and pseudo- and anti-pseudo-Hermiticity ($\scQ_+$ and $\scQ_-$). In the Bernard-LeClair classification scheme~\cite{bernard2002}, $\scT_\pm$ are referred to as K symmetries, $\scC_\pm$ as C symmetries, $\scP$ as P symmetry, and $\scQ_\pm$ as Q symmetries. Carefully accounting for all inequivalent combinations of independent symmetries, the values of the square of the antiunitary ones, and the commutation or anticommutation relations of the unitary involutions gives 38 non-Hermitian symmetry classes~\cite{bernard2002,magnea2008,zhou2019,kawabata2019} for point-gap spectra and 54 classes for line-gap spectra~\cite{liu2019PRB,ashida2020}. In the Hermitian case, the classification simplifies to the symmetries of Eqs.~(\ref{eq:nHsym_Tp}), (\ref{eq:nHsym_Cm}), and (\ref{eq:nHsym_P}), and leads to the tenfold classification of Altland and Zirnbauer~\cite{altland1997}.

\subsection{Lindbladians without unitary symmetries}

The spectrum of the Lindbladian cannot be freely rotated since there is a preferred axis of symmetry (the negative real axis), and hence Lindbladians belong to one of the 54 line-gap spectra classes. However, not all these symmetry classes can be realized in Lindbladian dynamics because of the special structure of the Lindblad superoperator. 

First, we notice that because $\scL$ preserves the Hermiticity of the density matrix, $\(\scL \rho\)^\dagger=\scL \rho^\dagger$, the eigenvalues of $\scL$ come in complex-conjugate pairs and we always have a $\scT_+$ symmetry squaring to $+1$, given by Eq.~(\ref{eq:nHsym_Tp}) with $\scT_+=\scK\swap$, where $\scK$ is the complex-conjugation superoperator defined by $\scK\rho=\rho^*$ and $\scK \scL \scK^{-1}=\scL^*$, and the \textsc{swap} operator $\swap$ exchanges the two copies of the doubled Hilbert space, $\swap \(A\otimes B\) \swap=B\otimes A$ for any operators $A,B$, and satisfies $\swap^2=+1$. Obviously, the same conclusion holds for the shifted Lindbladian $\scL'$. There are 15 symmetry classes out of the 54 that satisfy $\scT_+^2=+1$ (dubbed AI, AI$_+$, AI$_-$, BDI$^\dagger$, DIII$^\dagger$, BDI, CI, BDI$_{++}$, BDI$_{+-}$, BDI$_{-+}$, BDI$_{--}$, CI$_{+-}$, CI$_{++}$, CI$_{--}$, and CI$_{-+}$). Second, a transposition symmetry $\scC_+$ is also allowed and determines the bulk level repulsion~\cite{hamazaki2019}. 

By considering the bare Lindbladian $\scL$, it would seem we have exhausted the possible symmetries. Indeed, because $\scL$ is trace preserving and completely positive, its spectrum always has a zero eigenvalue (corresponding to the steady state) and the remaining eigenvalues have nonpositive real parts, which forbids any possible symmetries that reflect the spectrum across either the origin or the imaginary axis~\cite{lieu2018}, i.e., $\scT_-$ and $\scC_-$. On the other hand, the spectrum of the shifted Lindbladian $\scL'$ is centered at the origin and there are eigenvalues with both positive and negative real parts. Hence, both $\scT_-$ and $\scC_-$ are allowed symmetries of $\scL'$. This is an immediate consequence of the well-known fact that while the involutive symmetries are usually stated as in Eqs.~(\ref{eq:nHsym_Tp})--(\ref{eq:nHsym_Qm}), they need only hold up to addition of multiples of the identity. For instance, we can modify Eq.~(\ref{eq:nHsym_Tm}) to
\begin{equation}
\label{eq:nHsym_Tm_scL'}
\scT_- \scL \scT_-^{-1} = -\scL+2\alpha \scI,
\end{equation}
for some real constant $\alpha$. If we take $\alpha$ to be as defined in Eq.~(\ref{eq:scL_prime}), the previous equation can be rewritten as a standard symmetry condition for $\scL'$:
\begin{equation}
\scT_- \scL' \scT_-^{-1} = -\scL'.
\end{equation}
The $\scC_-$, $\scP$, and $\scQ_-$ symmetry transformations in Eqs.~(\ref{eq:nHsym_Cm}), (\ref{eq:nHsym_P}), and (\ref{eq:nHsym_Qm}), have to be redefined in the same way. On the other hand, no redefinition of $\scT_+$, $\scC_+$, and $\scQ_+$ symmetries is necessary, as we can trivially add $-\alpha\scI$ to both sides of Eqs.~(\ref{eq:nHsym_Tp}), (\ref{eq:nHsym_Cp}), and (\ref{eq:nHsym_Qp}) to rewrite them in terms of $\scL'$.
The possibility of shifting the spectrum is usually ignored because shifts in energy are irrelevant; i.e., we can always choose Hamiltonians to be traceless. However, the trace of the Lindbladian is not arbitrary and generalized transformations in terms of $\scL'$ have to be considered. 

Before proceeding, we note that instead of organizing the 15 classes in terms of the antiunitary symmetries present besides the $\scT_+$ symmetry, it will also prove convenient to alternatively label a class by its unitary involutions $\scP$ and $\scQ_\pm$. This also offers a check on our counting of the classes: there is one class with no unitary involutions; if there is one additional unitary involution, it can be either $\scP$, $\scQ_+$, or $\scQ_-$, and in each case it can either commute or anticommute with $\scT_+$, i.e., $3\times2=6$ classes; if two additional unitary involutions are present we can, without loss of generality, consider them to be $\scP$ and $\scQ_+$ (the other two combinations are obtained by taking one of $\scP$ or $\scQ_+$ and their product as the two independent involutions, since the product $\scP \scQ_+$ is a $\scQ_-$ symmetry), which either commute or anticommute with each other and with $\scT_+$, i.e., $2\times2\times2=8$ classes; there is no class with the three involutions since the product $\scP\scQ_+\scQ_-$ is a unitary symmetry commuting with the Lindbladian, which we assume not to exist; in total, we thus have $1+6+8+0=15$ classes. The $\scP$ and $\scQ_\pm$ symmetries of the Lindbladian then induce antiunitary symmetries through the relations
\begin{equation}\label{eq:antiunitary_from_unitary}
\scT_-=\scP \scT_+,
\quad
\scC_-=\scQ_- \scT_+,
\quad \text{and}\quad
\scC_+=\scQ_+ \scT_+.
\end{equation}
Furthermore, the square of the antiunitary symmetries and the commutation relations of the unitary involutions are related by
\begin{align}
\label{eq:sq_commT}
    &\scT_-^2=\epsilon_{\scP\!\scT_+}\scT_+^2,
    \\
    &\scC_\pm^2=\epsilon_{\scQ_\pm\!\scT_+}\scT_+^2
    =
    \label{eq:sq_commC}\epsilon_{\scP\!\scT_+}\epsilon_{\scQ_\mp\!\scT_+}\epsilon_{\scP\!\scQ_\mp}\scT_+^2.
\end{align}
These two ways of labeling symmetry classes are equivalent and are used interchangeably in what follows. In the remainder of this section and in Sec.~\ref{sec:examples}, we use the unitary involutions, while the discussion of random-matrix universality in Sec.~\ref{sec:rmt} is based on antiunitary symmetries.

One might be tempted to conclude that no further restrictions on the symmetries of $\scL'$ exist and, thus, that there are 15 symmetry classes of many-body Lindbladians. However, the Lindbladian is not an arbitrary superoperator with $\scT_+^2=1$ symmetry, and has an additional structure in terms of the Hamiltonian and jump operators. In Sec.~\ref{subsec:conditions}, we derive the conditions these operators must satisfy in order to implement a superoperator symmetry of the Lindbladian. Based on these conditions, in Sec.~\ref{subsec:noCp2-1} we argue that, remarkably, a $\scC_-^2=-1$ symmetry is not allowed. Since there are five classes out of the 15 (DIII$^\dagger$, BDI$_{+-}$, BDI$_{--}$, CI$_{++}$, and CI$_{-+}$) with $\scC_-^2=-1$, a Liouvillian without unitary symmetries belongs to one of ten non-Hermitian symmetry classes, which are listed in Table~\ref{tab:Lindbladian classes} together with their defining relations and matrix realizations. 

\begin{table*}[t]
    \caption{Non-Hermitian symmetry classes with $\scT_+^2=+1$ and $\scC_+^2\neq-1$, which can realized by Lindbladians with unbroken $\scT_+$ symmetry. For each class, we list its Bernard-LeClaire (BL) symmetries, the nomenclature following Ref.~\cite{kawabata2019}, the squares of its antiunitary symmetries, its unitary involutions and their commutation relations [as defined in Eqs.~(\ref{eq:eps_1})--(\ref{eq:eps_5})], and an explicit matrix realization. In the second column, we have adopted a shorthand notation, where the commutation relations of $\scP$ symmetry are indicated with a subscript in the class name (class AI$_+$, say, is denoted AI + $\mathcal{S}_+$ in Ref.~\cite{kawabata2019}). Moreover, these class names are not unique (for instance, class AI is also known as D$^\dagger$, and class BDI$_{-+}$ as CI$^\dagger_{+-}$, DIII$_{-+}$, or BDI$^\dagger_{+-}$~\cite{kawabata2019}). In the matrix realizations of the last column, $A$, $B$, $C$, and $D$ are arbitrary non-Hermitian matrices unless specified otherwise and empty entries correspond to zeros.}
	\label{tab:Lindbladian classes}
	\begin{tabular}{@{}Sl Sl Sc Sc Sc Sc Sc Sc Sc Sc Sc Sl@{}}
		\toprule
		BL symmetry & Class & $\scT_+^2$ & $\scC_-^2$ & $\scC_+^2$ & $\scT_-^2$ & $\epsilon_{\scP\!\scT_+}$ & $\epsilon_{\scQ_+\!\scT_+}$ & $\epsilon_{\scQ_-\!\scT_+}$ & $\epsilon_{\scP\!\scQ_+}$ & $\epsilon_{\scP\!\scQ_-}$ & Matrix realization  \\ \midrule
		1, K    & AI             & $+1$ & ---  & ---  & ---  & ---  & ---  & ---  & ---  & ---  & $A=A^*$       \\ \midrule
		2, PK   & AI$_+$         & $+1$ & ---  & ---  & $+1$ & $+1$ & ---  & ---  & ---  & ---  & $\matAIp$     \\ \midrule
		3, PK   & AI$_-$         & $+1$ & ---  & ---  & $-1$ & $-1$ & ---  & ---  & ---  & ---  & $\matAIm$     \\ \midrule
		4, QC   & BDI$^\dagger$  & $+1$ & ---  & $+1$ & ---  & ---  & $+1$ & ---  & ---  & ---  & $\matBDId$    \\ \midrule
		5, QC   & BDI            & $+1$ & $+1$ & ---  & ---  & ---  & ---  & $+1$ & ---  & ---  & $\matBDIalt$  \\ \midrule
		6, QC   & CI             & $+1$ & $-1$ & ---  & ---  & ---  & ---  & $-1$ & ---  & ---  & $\matCIalt$   \\ \midrule
		7, PQC  & BDI$_{++}$     & $+1$ & $+1$ & $+1$ & $+1$ & $+1$ & $+1$ & ---  & $+1$ & ---  & $\matBDIdpp$  \\ \midrule
            8, PQC & BDI$_{-+}$     & $+1$ & $+1$ & $+1$ & $-1$ & $-1$ & $+1$ & ---  & $-1$ & ---  & $\matBDIdpm$  \\ \midrule
            9, PQC & CI$_{+-}$      & $+1$ & $-1$ & $+1$ & $+1$ & $+1$ & $+1$ & ---  & $-1$ & ---  & $\matBDIdmp$  \\ \midrule
		10, PQC & CI$_{--}$      & $+1$ & $-1$ & $+1$ & $-1$ & $-1$ & $+1$ & ---  & $+1$ & ---  & $\matBDIdmm$ \\
        \bottomrule
	\end{tabular}
\end{table*}

\subsection{Lindbladians with unitary symmetries}
\label{subsec:unitary_syms}

Let us now consider the consequences of a unitary symmetry $\scU$ commuting with the Lindbladian. These symmetries come in two types~\cite{buca2012} (strong and weak). If the Hamiltonian and jump operators jointly satisfy the symmetry relations
\begin{align}
\label{eq:strong_symm}
    \comm{u}{H}=\comm{u}{L_m}=0, \quad m=1,\dots, M,
\end{align}
then both unitary superoperators
\begin{align}
    \scU_\mathrm{L}=u\otimes \id 
    \qquad \text{and} \qquad
    \scU_\mathrm{R}=\id \otimes u^*
\end{align}
commute with the Lindbladian and we refer to them as a Liouvillian strong symmetry~\cite{buca2012}. There are $n$ quantum numbers in \emph{each copy} in the doubled Liouville space $\mathcal{H}\otimes\mathcal{H}$ (where $n$ denotes the number of distinct eigenvalues of $u$), which are conserved independently. The Liouville space thus splits into $n^2$ invariant subspaces (symmetry sectors) and there are generically $n$ distinct steady states, one in each diagonal sector with equal quantum numbers in the two copies. 

If the relations of Eq.~(\ref{eq:strong_symm}) are not all satisfied simultaneously, but
\begin{equation}
    \scU=\scU_\mathrm{L}\scU_\mathrm{R}=u\otimes u^*
\end{equation} 
still commutes with $\scL'$, we call it a weak Liouvillian symmetry~\cite{buca2012}. There are between $n$ and $n(n-1)$ invariant subspaces (depending on the precise form of the symmetry $u$) and generically a single steady state (in the symmetry sector with eigenvalue $1$). For additional details, see Ref.~\cite{buca2012}.

We block diagonalize $\scL'$, such that each block has a well-defined eigenvalue of $\scU$ (weak symmetry) or $\scU_\mathrm{L}$, $\scU_\mathrm{R}$ (strong symmetry). For a given block to belong to a certain symmetry class, the antiunitary symmetries $\scT_\pm$ and $\scC_\pm$ and the unitary involutions $\scP$ and $\scQ_\pm$ defining that class must act within the block, i.e., they must commute with the projector onto that block. If they mix different blocks (because they connect eigenstates in different symmetry sectors), the superoperator symmetry is broken in those blocks, although the full Lindbladian possesses the symmetry.

Following the previous considerations, we immediately conclude that the presence of commuting unitary symmetries enriches the symmetry classification of the Lindbladian and allows us to go beyond the tenfold classification: if the $\scT_+=\scK\swap$ symmetry is broken by $\scU$ and no other independent $\scT_+$ symmetry is realized, the irreducible block of the Lindbladian belongs to one of 19 symmetry classes with no $\scT_+$ and $\scC_+^2\neq-1$. Note that the same arguments put forward in Sec.~\ref{subsec:noCp2-1} that prohibit a $\scC_+^2=-1$ symmetry also preclude a $\scT_+^2=-1$ symmetry. Moreover, in these classes, because of the absence of $\scT_+$ symmetry, the existence of a $\scP$ or $\scQ_\pm$ symmetry is independent of the existence of a $\scT_-$ or $\scC_\pm$ symmetry, respectively, and, therefore, a careful counting leads to 19 independent symmetry classes. Remarkably, the projection of the Lindbladian into a symmetry sector that contains a steady state (eigenvalue zero) always preserves the $\scT_+$ symmetry, as we discuss in detail below for the two cases of strong and weak symmetries. $\scT_+^2=+1$ symmetry is only broken in the blocks without the steady states (i.e., without the eigenvalue zero), which correspond to short-lived transient dynamics. We thus reach the conclusion that many-body Lindbladians admit a ($10+19$)-fold symmetry classification: in the absence of unitary symmetries \emph{or} in the presence of unitary symmetries in all symmetry sectors containing the steady state(s), the Lindbladian belongs to one of ten non-Hermitian symmetry classes with $\scT_+^2=+1$; if however, there are additional unitary symmetries and we consider non-steady-state sectors, the Lindbladian may belong to a different set of 19 classes with broken $\scT_+$ symmetry.

The simplest way to break $\scT_+$ symmetry occurs when $\scU$ and $\scT_+\scK$ do not commute, and hence do not share a common eigenbasis. As an example, we mention the case of a Liouvillian strong symmetry: since the transformation acts as a symmetry of each copy of Hilbert space individually, by definition it does not commute with the \textsc{swap} operator implementing the $\scT_+$ symmetry. Then, the $\scT_+$ is unbroken in the blocks with the same quantum number in both copies (the blocks containing the $n$ steady states) and is broken in the remaining.

However, even if $\scT_+\scK$ and $\scU$ commute, the $\scT_+$ symmetry can be broken if $\scT_+$ does not commute with the projector onto a specific subspace. Since $\scU$ has complex unimodular eigenvalues and the $\scT_+$ symmetry involves complex conjugation, states in the sector with quantum number $e^{i\theta}$ are transformed into states in the sector with quantum number $e^{-i\theta}$ under the action of $\scT_+$. The $\scT_+$ symmetry is preserved in the sectors with quantum number $\pm 1$ and broken in all others. The sector with quantum number $+1$ always exists (and contains the steady state)~\cite{buca2012}, while the additional $\scT_+$-unbroken sector with eigenvalue $-1$ might or might not exist (it does for a $\mathbb{Z}_2$ symmetry, which will be relevant below).

Before concluding this section, let us briefly comment on the impossibility of implementing a Lindbladian class with $\scC_+^2=-1$. Indeed, as we show in Sec.~\ref{subsec:noCp2-1}, the existence of a $\scC_+^2=-1$ symmetry always requires the presence of a strong symmetry. Moreover, we also show that, under quite general conditions, the $\scC_+^2=-1$ symmetry, when it exists, always breaks the strong symmetry it induces. Therefore, even when a $\scC_+^2=-1$ symmetry exists (with physical consequences such as Liouvillian or open system version of Kramers degeneracy~\cite{lieu2022}), it does not define a symmetry class with $\scC_+^2=-1$ (which has consequences, for instance, for level statistics). 
The same argument also implies that a $\scT_+^2=-1$ symmetry cannot exist. Hence, if a unitary symmetry breaks the $\scT_+=\scK\swap$ symmetry, an alternative $\scT_+^2=-1$ symmetry cannot be implemented, supporting our counting of 19 classes above.

In summary, many-body Lindbladians have a tenfold classification in the absence of unitary symmetries. The presence of the latter allows for 19 additional classes beyond the tenfold way. Since the Lindbladian is specified in terms of its Hamiltonian and jump operators, it is natural to ask what conditions these operators must satisfy for the Lindbladian superoperator to belong to one of the classes discussed above. We address this question in the next section.

\subsection{Conditions on the Hamiltonian and jump operators}
\label{subsec:conditions}

In this section, we derive sufficient operator conditions for inducing superoperator symmetries of the Lindbladian. We see that these conditions are different for the three contributions to the Lindbladian, $\scLH$, $\scLD$, and $\scLJ$. We state the conditions in terms of the unitary involutions $\scP$ and $\scQ_\pm$. As mentioned above, they could be alternatively expressed in terms of the antiunitary symmetries $\scT_-$ and $\scC_\pm$.

\subsubsection{Jump term}
\label{subsec:sym_scLJ}

To impose superoperator $\scP$ and $\scQ_\pm$ symmetries on $\scLJ$, we impose operator $\scP$ and $\scQ_\pm$ symmetries on the jump operators. Since we always work with traceless jump operators, we do not need to consider the symmetries of the shifted jump operators. Furthermore, we do not require that each $L_m$ transforms to itself under the symmetries, only that the complete set of jump operators is closed under it. More precisely, we consider $L_m$ that satisfy
\begin{alignat}{99}
\label{eq:symL_p}
&p_a L_m p_a^{-1} &&= \epsilon^L_{pa}\sum_{n=1}^M P_{mn} L_n,\qquad 
&&p_{a}^2=+1,
\\
\label{eq:symL_q}
&q_a L_m^\dagger q_a^{-1} &&= \epsilon^L_{qa} \sum_{n=1}^M Q_{mn} L_n,\qquad 
&&q_{a}^2=+1,
\end{alignat}
where $a=1,2$, $\epsilon^L_{pa},\epsilon^L_{qa}=\pm1$, $p_{a}$ and $q_{a}$ are unitary and Hermitian, and $P$ and $Q$ are $M\times M$ unitary Hermitian and unitary symmetric matrices, respectively.
Note that the index $a$ allows for more than one symmetry of each type, but that it is also possible that only one exists, in which case $p_1=p_2$ or $q_1=q_2$.
Next, we define the unitary superoperators:
\begin{alignat}{99}
\label{eq:sym_scL_Pab}
&\scP_{ab}=p_a\otimes p_b^*,\qquad 
&&\scP_{ab}^2=+1,
\\
\label{eq:sym_scL_Qab}
&\scQ_{ab}=q_a\otimes q_b^*,\qquad 
&&\scQ_{ab}^2=+1.
\end{alignat}
It is straightforward to check that $\scP_{ab}$ acts as either a commuting, unitary, or an anticommuting, chiral, symmetry of $\scLJ$ [defined in Eq.~(\ref{eq:scLJ})]:
\begin{equation}
\label{eq:P_LJ_rel}
\begin{split}
\scP_{ab}\scLJ\scP_{ab}^{-1}
&=2\sum_{m}\(p_a\otimes p_{b}^*\)\(L_m\otimes L_m^*\) \(p_a \otimes p_{b}^*\)
\\
&=2\epsilon^L_{pa}\epsilon^L_{pb}\sum_{mnp} P_{mn} P_{mp}^*\, L_n \otimes L_p^*
\\
&=\epsilon^L_{pa}\epsilon^L_{pb}\,\scLJ,
\end{split}
\end{equation} 
where we use Eq.~(\ref{eq:symL_p}) and the unitarity of $P$. If $\epsilon^L_{pa}=\epsilon^L_{pb}$, $\scP_{ab}$ acts as a commuting unitary symmetry of $\scLJ$; if $\epsilon^L_{pa}=-\epsilon^L_{pb}$, it acts as a chiral symmetry.  

Similarly, we can show that $\scQ_{ab}$ acts as a $\scQ_\pm$ symmetry of $\scLJ$, depending on the signs $\epsilon^L_{qa}$, $\epsilon^L_{qb}$: if $\epsilon^L_{qa}=\epsilon^L_{qb}$, $\scQ_{ab}$ acts as a $\scQ_+$ symmetry; if $\epsilon^L_{qa}=-\epsilon^L_{qb}$, as a $\scQ_-$ symmetry.

\subsubsection{Dissipative term}
\label{subsec:sym_scLD}

The conditions of Eq.~(\ref{eq:symL_p}) and (\ref{eq:symL_q}) are not enough to generate symmetries of $\scLD$. For instance, a chiral symmetry $p_a$, Eq.~(\ref{eq:symL_p}), does not modify the term $\sum_m L_m^\dagger L_m$:
\begin{equation}
\begin{split}
p_a \(\sum_{m} L_m^\dagger L_m\) p_a^{-1}
&=\sum_m p_a L_m^\dagger p_a^{-1}p_aL_m p_a^{-1}
\\
&=+\sum_m L_m^\dagger L_m,
\end{split}
\end{equation}
and, hence,
\begin{equation}
\label{eq:transf_scP_scLD}
\scP_{ab}\scLD \scP_{ab}^{-1}=+\scLD;
\end{equation}
i.e., the condition of Eq.~(\ref{eq:symL_p}) only leads to a commuting unitary symmetry of $\scLD$, not to a chiral symmetry. To generate a $\scP$ symmetry, we have to require that the jump operators additionally satisfy
\begin{equation}\label{eq:sym_scLJ_P}
\sum_m L_m^\dagger L_m = -\frac{\alpha}{2} \id,
\end{equation}
where $\alpha$ is defined in Eq.~(\ref{eq:scL_prime}). In that case, Eq.~(\ref{eq:transf_scP_scLD}) reads as
\begin{equation}\label{eq:scTab_scLD}
\scP_{ab}\scLD \scP_{ab}^{-1}
=-\scLD+2\alpha \scI
\iff
\scP_{ab}\scLD'\scP_{ab}^{-1}=-\scLD',
\end{equation}
in accordance with Eq.~(\ref{eq:nHsym_Tm_scL'}). 
Note that, for this particular symmetry only, we actually have $\scLD'=0$.
We thus see that it is the dissipative contribution that forces us to consider the symmetries of the shifted Lindbladian. One particular way of satisfying Eq.~(\ref{eq:sym_scLJ_P}), which we encounter in the examples below, is to have each jump operator individually satisfy
\begin{equation}\label{eq:sym_scLJ_P_m}
L_m^\dagger L_m = -\frac{\alpha_{m}}{2} \id,
\end{equation}
for some $\alpha_m\in\mathbb{R}$.

Similarly, we can see that a pseudo-Hermiticity transformation, Eq.~(\ref{eq:symL_q}), transforms the term $\sum_m L_m^\dagger L_m$ as
\begin{equation}
q_a\(\sum_m L_m^\dagger L_m\)^\dagger q_a^{-1}=\sum_{m}L_mL_m^\dagger,
\end{equation}
and hence we must impose a condition on the commutator or anticommutator of $L_m$. If we impose that
\begin{equation}\label{eq:sym_scLJ_Q-}
\sum_m\acomm{L_m^\dagger}{L_m}=-\alpha \id,
\end{equation}
then $\scQ_{ab}$ acts as a $\scQ_-$ symmetry:
\begin{equation}
\scQ_{ab}\scLD^\dagger \scQ_{ab}^{-1}
=-\scLD+2\alpha \scI
\iff
\scQ_{ab}\scLD'^\dagger\scQ_{ab}^{-1}=-\scLD'.
\end{equation}
Again, it will often prove convenient for each jump operator to satisfy this condition individually; i.e.,
\begin{equation}\label{eq:sym_scLJ_Q-_m}
\acomm{L_m^\dagger}{L_m}=-\alpha'_m \id,
\end{equation}
for some $\alpha'_m\in\mathbb{R}$. 
If we instead impose that
\begin{equation}\label{eq:sym_scLJ_Q+}
\sum_m\comm{L_m^\dagger}{L_m}=0
\end{equation}
(or $\comm{L_m}{L_m^\dagger}=0$ for each jump operator), then $\scQ_{ab}$ acts as a $\scQ_+$ symmetry, $\scQ_{ab}\scLD'^\dagger \scQ_{ab}^{-1}=\scLD'$. 

\subsubsection{Hamiltonian term}
\label{subsec:sym_scLH}

Finally, we address the conditions one has to impose on the Hamiltonian such that $\scLH$ possesses $\scP$ and $\scQ_\pm$ symmetries. We start from the symmetries of the Hamiltonian:
\begin{align}
\label{eq:symH}
&p_a H p_a^{-1} = \epsilon^H_{pa} H,
\\
&q_a H q_a^{-1} = \epsilon^H_{qa} H.
\end{align}
Note that for the full Lindbladian to satisfy a superoperator symmetry, the matrices $p_a$ and $q_a$ have to be the same as those in Eqs.~(\ref{eq:symL_p}) and (\ref{eq:symL_q}). Under the action of the superoperator $\scP_{ab}$, $\scLH$ transforms as
\begin{equation}
\begin{split}
\scP_{ab} \scLH \scP_{ab}^{-1} 
&=-\i \(p_a H p_a^{-1}\otimes \id - \id \otimes \(p_b H p_b^{-1}\)^*\)
\\
&=-\i \(\epsilon^H_{pa}H\otimes\id-\epsilon^{H}_{pb}\id\otimes H^*\).
\end{split}
\end{equation}
For $\scP_{ab}$ to be a symmetry of $\scLH$, we must have $\epsilon^H_{pa}=\epsilon^H_{pb}$. Then, if $\epsilon^H_{pa}=\epsilon^H_{pb}=-1$, we have
\begin{equation}
\scP_{ab} \scLH \scP_{ab}^{-1} 
=\i \(H\otimes\id-\id\otimes H^*\)
=-\scLH,
\end{equation}
i.e., $\scP_{ab}$ acts as a $\scP$ symmetry, while if $\epsilon^H_{pa}=\epsilon^H_{pb}=+1$, it acts as a commuting unitary symmetry.

Proceeding analogously for the pseudo-Hermiticity transformations, we find $\epsilon^H_{qa}=\epsilon^H_{qb}$. If $\epsilon^H_{qa}=\epsilon^H_{qb}=-1$, then $\scQ_{ab}$ acts as $\scQ_+$ symmetry and if $\epsilon^H_{qa}=\epsilon^H_{qb}=+1$, it acts as a $\scQ_-$ symmetry. 

Finally, we note that in the case of real Hamiltonian and jump operators, we can define a modified superoperator $\widetilde{\scQ}_{ab}=\scQ_{ab}\swap$ whose action on $\scLH$ is reversed: if $\epsilon^H_{qa}=\epsilon^H_{qb}=+1$, $\widetilde{\scQ}_{ab}$ is a $\scQ_+$ symmetry, while it is a $\scQ_-$ symmetry if $\epsilon^H_{qa}=\epsilon^H_{qb}=-1$. Similarly, if the Hamiltonian is real and the jump operators are symmetric, we can define $\widetilde{\scP}_{ab}=\scP_{ab}\swap$, such that if $\epsilon^H_{pa}=\epsilon^H_{pb}=+1$, $\widetilde{\scP}_{ab}$ is a $\scP$ symmetry of $\scLH$, while it is a commuting unitary symmetry when $\epsilon^H_{pa}=\epsilon^H_{pb}=-1$. In either case, the action on $\scLJ$ and $\scLD$ is not modified.

\subsection{Absence of $\scC_+^2=-1$ symmetry and Kramers degeneracy}
\label{subsec:noCp2-1}

We now show that, under fairly general conditions, classes with $\scC_+^2=-1$ do not exist in the Lindbladian classification. The proof proceeds in two steps. First, we show that, because of the two-copy tensor-product structure of the Lindbladian, a $\scC_+^2=-1$ symmetry always implies the existence of a Liouvillian strong symmetry. Then, we show that, by construction, the $\scC_+^2=-1$ is always broken by the strong symmetry it induces. As a consequence, if degenerate Kramers pairs exist, they do not occur in the same symmetry sector, and none of the blocks of the Lindbladian displays, by itself, Kramers degeneracy. 
Importantly, the absence of Kramers pairs inside individual symmetry sectors is a rather universal result of systems with a two-copy structure and symmetric intercopy coupling, as it is also observed for fermionic Lindbladians~\cite{kawabata2022Classes} and for a Hermitian two-site fermionic Sachdev-Ye-Kitaev Hamiltonian~\cite{sa2023}, where an identical argument holds.
The same mechanism also prevents the existence of a $\scT_+^2=-1$ symmetry, which does not affect the ten classes with unbroken $\scT_+$ swap symmetry, but it is fundamental in the counting of the 19 classes with broken $\scT_+$ symmetry (as it precludes any additional classes with $\scT_+^2=-1$).

The first part of the proof is completely general. Let us assume that a superoperator symmetry $\scQ_+=q_a\otimes q_b^*$ of the Lindbladian exists. From Eq.~(\ref{eq:sq_commC}), we have that $\scC_+^2=\epsilon_{\scQ_+\!\scT_+}$. The commutation relation of $\scQ_+$ with $\scT_+$ is given by:
\begin{equation}
\begin{split}
    \scQ_+\scT_+&=(q_a\otimes q_b^*)\scK\swap 
    \\&=\scK \swap (q_b\otimes q_a^*)
    \\&=\scT_+ \scQ_+ \left[q_a^{-1}q_b\otimes (q_b^{-1}q_a)^* \right],
\end{split}
\end{equation}
We want to impose that $\scC_+^2=-1\Leftrightarrow\scQ_+\scT_+=-\scT_+\scQ_+$. Clearly, that is not possible if $q_a=q_b$, i.e., if the Hamiltonian and jump operators have a single $\scQ_+$ operator symmetry. Consequently, we must consider a $\scQ_+$ symmetry of the form
\begin{equation}
    \scQ_+=q_1\otimes q_2^*,
\end{equation}
with $q_1\neq q_2$. Furthermore, to implement the unitary involution $\scQ_+$, the Hamiltonian and jump operators must satisfy $\epsilon_{q1}^H=\epsilon_{q2}^H=-1$ and $\epsilon_{q1}^L=\epsilon_{q2}^L$, according to the previous section. The former condition further precludes that one of the $q_{1,2}$ is the identity operator. It then immediately follows that the product $q_1q_2$ commutes with the Hamiltonian and all jump operators and thus implements a Liouvillian strong symmetry.

To conclude the proof, we must show that if the $\scC_+^2=-1$ symmetry exists, it is always broken by the strong symmetry it induces. Let us define matrices $\varepsilon_{12}$, $\eta_1$, and $\eta_{2}$ through the relations
\begin{equation}
\label{eq:Kramers_assump}
    q_1 q_2=\varepsilon_{12}q_2 q_1
    \qquad \text{and} \qquad
    q_{1,2}=\eta_{1,2} q_{1,2}^*.
\end{equation}
Because $q_{1,2}$ are unitary, it immediately follows that $\varepsilon_{12}$ and $\eta_{1,2}$ must also be unitary. To proceed, we make the mild assumption that $\varepsilon_{12}$ and $\eta_{1,2}$ are unimodular complex numbers (i.e., proportional to the identity).
This assumption holds for any $q_1$ and $q_2$ that can be expressed as a string of Pauli operators (which is true for the spin-chain examples of Sec.~\ref{sec:examples} and for fermionic models not discussed in this paper~\cite{garcia2022PRX,kawabata2022Classes,sa2023}). While the proof we present in the following strictly holds only in this case, we believe the argument extends to general $q_a$ written as sums of such Pauli strings (for which $\varepsilon_{12}$ and $\eta_{1,2}$ are more general unitary matrices) and, consequently, that sectors with $\scC_+^2=-1$ do not exist in general.

Proceeding under the assumption that $\varepsilon_{12}$ is a complex unimodular number, we take the strong symmetry to be implemented by the unitary
\begin{equation}
    u=q_1 q_2,
\end{equation}
which satisfies $u^2=\varepsilon_{12}$ and $u^*=\eta_1\eta_2u$.
Since $u$ defines a strong symmetry, both 
\begin{equation}
    \scU_\mathrm{L} = u\otimes \id 
    \quad \text{and} \quad
    \scU_\mathrm{R} = \id \otimes u^* = \eta_1 \eta_2 \(\id\otimes u\)
\end{equation}
are independently conserved, with eigenvalues $p_\mathrm{L,R}=\varepsilon_{12}^{1/2}$. The projectors onto the conserved sectors are 
\begin{equation}
    \mathbb{P}_{\mathrm{L},\mathrm{R}}^\pm 
    =\frac{1}{2}\(\scI \pm \scU_{\mathrm{L},\mathrm{R}}/p_{\mathrm{L},\mathrm{R}}\)
    =\frac{1}{2}\(\scI \pm \varepsilon_{12}^{-1/2}\scU_{\mathrm{L},\mathrm{R}}\).
\end{equation}
We can now start imposing conditions on the choice of operators. The $\scC_+^2$ symmetry squares to
\begin{equation}
\begin{split}
    \scC_+^2=\(\scQ_+\scT_+\)^2
    &= (q_1\otimes q_2^*)(q_2\otimes q_1^*)
    \\
    &= \varepsilon_{12} \(q_1 q_2 \otimes q_1^* q_2^*\)
    \\
    &= \varepsilon_{12}\scU_\mathrm{L}\scU_\mathrm{R}.
\end{split}
\end{equation}
Since we want $\scC_+^2=-1$, we must be in a sector of $\scU_{\mathrm{L},\mathrm{R}}$ with quantum numbers $p_\mathrm{L}p_\mathrm{R}\varepsilon_{12}=-1$. On the other hand, for $\scC_+$ to act inside a given symmetry sector of $u$, it must commute with the projector. The commutation relation is given by
\begin{equation}
\begin{split}
    \scC_+ \mathbb{P}_{\mathrm{L},\mathrm{R}}^\pm
    &= \scQ_+ \scT_+\mathbb{P}_{\mathrm{L},\mathrm{R}}^\pm
    \\
    &= (q_1\otimes q_2^*) \mathbb{P}_{\mathrm{R},\mathrm{L}}^{\pm\varepsilon_{12}} \scT_+
    \\
    &= \frac{1}{2}\(\scI\pm \varepsilon_{12}^{1/2}\scU_{\mathrm{L},\mathrm{R}}\)\scQ_+\scT_+
    \\
    &= \mathbb{P}_{\mathrm{L},\mathrm{R}}^{\pm p_\mathrm{L}p_\mathrm{R}\varepsilon_{12}}\scC_+.
\end{split}
\end{equation}
From this, it follows that $p_\mathrm{L}p_\mathrm{R}\varepsilon_{12}=+1$, in contradiction with the condition we found above. We conclude that either $\scC_+$ acts inside a sector but squares to $+1$, or it squares to $-1$ but connects different sectors. In either case, a definite symmetry sector does not belong to a class with $\scC_+^2=-1$.

As noted in the previous section, one might define an alternative symmetry operator $\scQ_+=(q_1\otimes q_2^*)\swap$. The calculation proceeds in the same way as above, and we find again two contradicting conditions: to have $\scC_+^2=-1$ we require $\eta_1\eta_2=-1$, while for $\scC_+$ to act inside a single symmetry sector, we must have $\eta_1\eta_2=+1$.

To conclude this section, note that the existence of $\scT_-^2=-1$ or $\scC_-^2=-1$ symmetries does not imply a strong symmetry because the jump operators must satisfy $\epsilon_{q1}^L\neq \epsilon_{q2}^L$ and, consequently, the product $q_1q_2$ anticommutes with the jump operators and can lead, at most, to a weak symmetry. On the other hand, if the $\scT_+=\scK\swap$ symmetry is broken, the same argument prevents the implementation of an alternative $\scT_+^2=-1$ symmetry. Hence, Lindbladian symmetry classes have either $\scT_+^2=+1$ or no $\scT_+$.

\subsection{Generalization to non-Markovian and non-trace-preserving open quantum dynamics}

After the developments of the previous sections, we are now in the position to make the remarkable observation that the classification we have developed is not restricted to Lindbladian dynamics, but to all Hermiticity-preserving dynamics, including non-Markovian and even non-trace-preserving dynamics.

To see this, we start from the fact that the Liouvillian generator $\Lambda$ of any Hermiticity-preserving quantum master equation, $\partial_t\rho=\Lambda \rho$, can be written in the form~\cite{hall2014PRA}
\begin{equation}
\label{eq:general_generator}
    \Lambda\rho 
    =
    -\i \comm{H}{\rho}+\acomm{\Gamma}{\rho}
    +2\sum_{m=1}^M \gamma_m L_m\rho L_m^\dagger,
\end{equation}
where, in addition to the Hamiltonian and jump operators, we have a second independent ``Hamiltonian'' $\Gamma=\Gamma^\dagger$, and the real rates $\gamma_m$ can be negative in general. Furthermore, we could also assume all of $H$, $L_m$, $\gamma_m$, and $\Gamma$ to be time dependent. 
In the most general case, the master equation~(\ref{eq:general_generator}), while Hermiticity preserving, is not necessarily positivity preserving~\cite{hall2014PRA}.
Trace preservation, $\partial_t\Tr\rho=\Tr\partial_t\rho=0$, is enforced by the restriction
\begin{equation}
    \Gamma=\sum_{m=1}^M \gamma_m L_m^\dagger L_m.
\end{equation}
Additionally, Markovianity is implemented by considering only positive rates $\gamma_m>0$. Then (and only then) they can be absorbed into the jump operators, $L_m\to \sqrt{\gamma_m}L_m$, and we recover the Liouvillian of Lindblad 
form (\ref{lindf}), $\Lambda=\scL$.

As before, the Liouvillian $\Lambda$ can be vectorized as $\Lambda=\Lambda_\mathrm{H}+\Lambda_\mathrm{D}+\Lambda_\mathrm{J}$, where the Hamiltonian, $\Lambda_\mathrm{H}$, and jump, $\Lambda_\mathrm{J}$, contributions are still given by Eqs.~(\ref{eq:scLH}) and (\ref{eq:scLJ}), respectively (apart from the real scalar rates $\gamma_m$ that do not change the classification), while the dissipative contribution $\Lambda_\mathrm{D}$ is now given by
\begin{equation}
    \Lambda_\mathrm{D}=\Gamma\otimes \id +\id \otimes \Gamma^*.
\end{equation}

It is now immediately clear that the classification (or, more precisely, the set of admissible classes) is not changed in this more general case. There is always a $\scT_+^2=1$ symmetry implemented by the \textsc{swap} operator (which can be broken by a Liouvillian strong symmetry), while the impossibility of $\scC_+^2=-1$ and $\scT_+^2=-1$ symmetries is imposed by the jump contribution and is, hence, unchanged. We thus have ten classes with unbroken $\scT_+$ symmetry and 19 additional ones with broken $\scT_+$ symmetry. 

What does change is the class to which a particular physical example is assigned. On the one hand, the jump operators no longer need to satisfy the strict conditions of Sec.~\ref{subsec:sym_scLD} [Eqs.~(\ref{eq:sym_scLJ_P}), (\ref{eq:sym_scLJ_Q-}), and (\ref{eq:sym_scLJ_Q+})], which facilitates finding examples in the classes with more symmetries. On the other hand, we have to impose constraints on the matrix $\Gamma$. More specifically, we consider that it admits the following symmetries:
\begin{align}
    p_a \Gamma p_a^{-1}=\epsilon_{pa}^\Gamma \Gamma,
    \\
    q_a \Gamma q_a^{-1}=\epsilon_{qa}^\Gamma \Gamma,
\end{align}
where the unitary operators $p_a$ and $q_a$ are the same as those in Secs.~\ref{subsec:sym_scLJ} and \ref{subsec:sym_scLH}, but the signs $\epsilon_{pa}^\Gamma$ and $\epsilon_{qa}^\Gamma$ are independent from the ones in the Hamiltonian and the jump contribution. Now, following exactly the same steps as in Sec.~\ref{subsec:sym_scLH} but noting that there is a factor-of-$\i$ difference between how $H$ and $\Gamma$ appear in the Liouvillian, we conclude that (i) $\scP_{ab}$ [defined in Eq.~(\ref{eq:sym_scL_Pab})] acts as a $\scP$ symmetry if $\epsilon_{pa}^\Gamma=\epsilon_{pb}^\Gamma=-1$ and as a commuting unitary symmetry if $\epsilon_{pa}^\Gamma=\epsilon_{pb}^\Gamma=+1$; (ii) $\scQ_{ab}$ [defined in Eq.~(\ref{eq:sym_scL_Qab})] acts as a $\scQ_+$ symmetry if $\epsilon_{qa}^\Gamma=\epsilon_{qb}^\Gamma=+1$ and as a $\scQ_-$ symmetry if $\epsilon_{qa}^\Gamma=\epsilon_{qb}^\Gamma=-1$; (iii) if $\Gamma$ is real symmetric, then we can define the alternative symmetry superoperators $\widetilde{\scP}_{ab}=\scP_{ab}\swap$ and $\widetilde{\scQ}_{ab}=\scP_{ab}\swap$, as discussed in Sec.~\ref{subsec:sym_scLH}, with the conditions on $\epsilon_{qa}^\Gamma, \epsilon_{qb}^\Gamma$ unchanged.

Since Hermiticity preservation is a physical constraint that one can hardly imagine to be relaxed, we conclude that our framework provides the most general symmetry classification of the dynamical generators of open quantum matter.

\subsection{Physical consequences for correlation functions}
\label{subsec:physical_consequences}

Before proceeding with specific examples of the classification developed so far, we derive general statements about the dynamics of open quantum systems described by any Lindbladian (or more general Liouvillian) with involutive global symmetries. Most importantly, we show that when the involutive symmetry involves a minus sign ($\scP$ or $\scQ_-$) we can derive a time-reversal-like invariance property for an observable in a time-dependent state, or a related correlation function.

We start with the case of a $\scP$ (or $\scT_-$) symmetry. For any fixed observable $O$ and state $\rho$ that are invariant under the $\scP$ operation, i.e., that satisfy the properties $\scP O=O$ and $\scP \rho=\rho$, we define the nonequilibrium correlation function
\begin{equation}
    F(t)=\Tr[O \rho(t)],
\end{equation}
where $\rho(t)=\exp{\scL t}\rho$ is the state evolved under the Lindbladian $\scL$ for time $t$. If $\scL$ satisfies Eq.~(\ref{eq:nHsym_P}), it follows that
\begin{equation}
    F(t)
    =e^{-2\alpha t}\Tr\left[O \scP e^{-\scL t} \scP \rho\right]
    =e^{-2\alpha t}\Tr\left[O \rho(-t)\right],
\end{equation}
or, equivalently,
\begin{equation}
\label{eq:correlation_Tm}
    F(-t)=e^{2\alpha t}F(t).
\end{equation}
For general open quantum systems, the quantity $F(-t)$ is not well defined: $-\scL$ does not generate a completely positive semigroup and, given a state at time $t$, we can only propagate it forward in time, not backward. The remarkable relation~(\ref{eq:correlation_Tm}) tells us, however, that in systems with a $\scP$ symmetry, $F(-t)$ is written in terms of two well-defined quantities [$F(t)$ and $\exp{2\alpha t}$] and is thus itself well defined. This opens the possibility of knowing the past of a dissipative system solely from the knowledge of its future. In particular, this feature could improve error-canceling schemes on noisy intermediate-scale quantum devices in combination with the recent proposal of Ref.~\cite{Minev}.

Next we consider $\scQ_\pm$ (equivalently, $\scC_\pm$) symmetries. Because these symmetries relate $\scL$ to its adjoint $\scL^\dagger$, we must consider correlation functions of two observables, or fidelity-like correlation functions of two states, $\rho$ and $\sigma$. Focusing on the latter case, we define
\begin{equation}
    G_{\rho\sigma}(t)=\Tr[\sigma \rho(t)],
\end{equation}
and consider states that are themselves invariant under the symmetry transformation, $\scQ_\pm \rho=\rho$ and $\scQ_\pm \sigma=\sigma$. If $\scL$ has a $\scQ_-$ symmetry, Eq.~(\ref{eq:nHsym_Qm}), we find, proceeding as before, that
\begin{equation}
    G_{\rho\sigma}(-t)=e^{2\alpha t} G_{\sigma \rho }(t),
\end{equation}
which, besides reversing time also swaps the two states. If, instead, the Lindbladian has a $\scQ_+$ symmetry, Eq.~(\ref{eq:nHsym_Qp}), no time reversal takes place and
\begin{equation}
    G_{\rho\sigma}(t)=G_{\sigma\rho}(t);
\end{equation}
i.e., $G$ is symmetric under the exchange of the two states $\sigma$ and $\rho$.

\section{Physical examples: Tenfold way in dissipative spin chains}
\label{sec:examples}

In the following sections, we realize the tenfold way of many-body Lindbladians with unbroken $\scT_+$ symmetry in spatially inhomogeneous spin chains. In Sec.~\ref{subsec:examples_IncHop}, we also present an example with a strong symmetry and, hence, sectors with broken $\scT_+$ symmetry. Throughout, we consider chains of $L$ spins $1/2$, represented by local Pauli operators $\sigma^{\alpha}_j=\id_{2\times 2}^{\otimes (j-1)}\otimes\sigma^\alpha\otimes \id_{2\times 2}^{\otimes(L-j)}$, $\alpha=x,y,z$, $j=1,2,\dots,L$, with periodic boundary conditions $\sigma_{L+1}^\alpha\equiv \sigma^\alpha_1$. We realize all ten symmetry classes by considering simple jump operators routinely used in the literature (dephasing, incoherent hopping, and spin injection or removal) and choosing an appropriate Hamiltonian. We thus conclude that the symmetry classes discuss in this work are not an exotic theoretical artifact, but are ubiquitous and implementable in current experimental setups.

\subsection{Dephasing. Classes BDI$_{++}$, CI$_{+-}$, BDI$_{-+}$, CI$_{--}$, BDI$^\dagger$, and AI}
\label{subsec:examples_Deph}

As a first example, we consider local dephasing jump operators,
\begin{equation}
L_j=\sqrt{\gamma_j}\sigma_j^z,
\end{equation}
where $\gamma_j$ are arbitrary positive dephasing rates. The trace of the Lindbladian is $\alpha=-2\sum_{j}\gamma_j$ and the shifted Lindbladian reads as
\begin{equation}
\scL'=-\i H\otimes \id +\id \otimes \i H+2\sum_{j=1}^L \gamma_j \sigma_j^z\otimes \sigma_j^z.
\end{equation}
Introducing the global spin operators,
\begin{equation}
\Sigma^\alpha=\prod_{j=1}^L \sigma^\alpha_j = (\sigma^\alpha)^{\otimes L},
\qquad
(\Sigma^\alpha)^2=+\id,
\end{equation} 
for $\alpha=x,y,z$, we can immediately check that the jump operators satisfy
\begin{align}
\label{eq:ex_deph_L_Sigma_comm}
\acomm{L_j}{\Sigma^x}=\acomm{L_j}{\Sigma^y}=\comm{L_j}{\Sigma^z}=0.
\end{align}
The dephasing Lindbladian is extremely rich, as the jump operators are real, Hermitian, and unitary. They thus satisfy all the conditions for symmetries of $\scLD$, Eqs.~(\ref{eq:sym_scLJ_P}), (\ref{eq:sym_scLJ_Q-}), and (\ref{eq:sym_scLJ_Q+}), allowing for the implementation of all three types of symmetries $\scP$ and $\scQ_\pm$ and realizing many different symmetry classes, depending on the choice of Hamiltonian. 

First, we consider a transverse-field Hamiltonian with a time-reversal-breaking interaction (not restricted to nearest neighbors):
\begin{equation}\label{eq:H_deph_noTRS}
    H=\sum_{j=1}^L g^x_j \sigma^x_j+\sum_{j<k}K_{jk}\sigma^y_j\sigma^z_k,
\end{equation}
with $g_j^x$ and $K_{jk}$ arbitrary real coupling constants.
The Hamiltonian satisfies
\begin{align}
\label{eq:ex_deph_H_Sigma_comm}
\comm{H}{\Sigma^x}=\acomm{H}{\Sigma^y}=\acomm{H}{\Sigma^z}=0.
\end{align}
From Eqs.~(\ref{eq:ex_deph_L_Sigma_comm}) and (\ref{eq:ex_deph_H_Sigma_comm}), it follows that the Lindbladian admits the commuting unitary symmetry (weak Liouvillian symmetry):
\begin{equation}\label{eq:ex_deph_scU}
\scU^x=\Sigma^x\otimes \Sigma^x,
\end{equation}
with eigenvalues $\pm1$. Accordingly, the Liouville space $(\mathbb C^2)^{\otimes L} \otimes (\mathbb C^2)^{\otimes L}$ splits into two sectors of positive or negative transverse parity ($\scU^x=\pm\scI$).
Moreover, Eqs.~(\ref{eq:ex_deph_L_Sigma_comm}) and (\ref{eq:ex_deph_H_Sigma_comm}) imply that 
\begin{equation}\label{eq:anti_ops_deph}
 \scP=\Sigma^z\otimes\Sigma^y 
 \quad\text{and}\quad
 \scQ_+=\Sigma^z\otimes \Sigma^z
\end{equation}
act as chiral symmetry and pseudo-Hermiticity of both the jump and Hamiltonian contributions. Both these symmetries and $\scT_+=\scK\swap$ commute with $\scU^x$ and, hence, act within the irreducible blocks of the Lindbladian. To identify the symmetry class of the (shifted) Lindbladian, we check the commutation relations of the $\scP$ and $\scQ_+$ operators:
\begin{align}
&\scP \scT_+ =(-1)^L\scU^x\, \scT_+ \scP,
\\
&\scQ_+ \scT_+ = \scT_+ \scQ_+,
\\
&\scQ_+ \scP =(-1)^L\, \scP \scQ_+.
\end{align}
Depending on the chain length and the parity sector, the Lindbladian belongs to different classes: for even $L$ and even parity ($\scU^x=+\scI$), it belongs to class BDI$_{++}$ (recall Table~\ref{tab:Lindbladian classes}); for even $L$ and odd parity ($\scU^x=-\scI$), to class CI$_{--}$; for odd $L$ and even parity, to class BDI$_{-+}$; and for odd $L$ and odd parity, to class CI$_{+-}$.
We note that the same symmetry classification holds if we add a second set of ``dephasing'' operators $\tilde{L}_j=\sqrt{\tilde{\gamma}_j}\sigma^y_j$.

As a second example, we choose a generic, time-reversal invariant XYZ Hamiltonian in a transverse field,
\begin{equation}
\label{eq:H_XYZ_X}
H=H_\mathrm{XYZ}+\sum_{j=1}^L g^x_j \sigma^x_j,
\end{equation}
with
\begin{equation}\label{eq:H_YXZ}
H_\mathrm{XYZ}=\sum_{j<k}
J_{jk}^x \sigma_j^x \sigma_k^x
+J_{jk}^y \sigma_j^y \sigma_k^y
+J_{jk}^z \sigma_j^z \sigma_k^z,
\end{equation}
and $J_{jk}^\alpha$ arbitrary real coupling constants. The Hamiltonian again commutes with $\Sigma^x$, but the anticommutation relations with $\Sigma^y$ and $\Sigma^z$ are broken. As before, the Lindbladian admits $\scU^x$ [Eq.(\ref{eq:ex_deph_scU})] as a weak Liouvillian symmetry. Because the Hamiltonian is real and the jump operators are real and symmetric, 
\begin{equation}
    \scP=\sqrt{p_x}(\Sigma^x\otimes \id)\swap
    \quad\text{and}\quad
    \scQ_+=\swap
\end{equation}
act as chiral symmetry and pseudo-Hermiticity of the jump and Hamiltonian contributions. Here, $p_x$ denotes the transverse parity, i.e., the eigenvalue of $\scU^x$, and is introduced in the definition of $\scP$ to ensure that $\scP^2=+1$ in both symmetry sectors. Both these symmetries and $\scT_+$ commute with $\scU^x$ and satisfy
\begin{align}
&\scP \scT_+ =\scT_+ \scP,
\\
\label{eq:ex_deph_comm_TQ}
&\scQ_+ \scT_+ =\scT_+ \scQ_+,
\\
&\scQ_+ \scP =\scU^x\, \scP \scQ_+.
\end{align}
The different parity sectors of the Lindbladian belong to different symmetry classes, this time irrespective of the chain length: in the sector of even parity ($\scU^x=+\scI$), the Lindbladian belongs to class BDI$_{++}$, whereas in the sector of odd parity ($\scU^x=-\scI$), it belongs to class CI$_{+-}$.

If we assume a more general Hamiltonian, we reduce the set of symmetries of the Lindbladian. If we add a second transverse component of the magnetic field say,
\begin{equation}
    H=H_\mathrm{XYZ}+\sum_{j=1}^L \left(g_j^x\sigma_j^x+h_j \sigma_j^z\right),
\end{equation}
we break the transverse parity conservation and all the commutation relations of the Hamiltonian. Consequently, the Lindbladian is irreducible and chiral symmetry $\scP$ is broken. $\scQ_+=\swap$ still implements a pseudo-Hermiticity transformation commuting with $\scT_+$ and, hence, the Lindbladian belongs to class BDI$^\dagger$.

Adding a third component to the magnetic field, i.e., setting
\begin{equation}
    H=H_\mathrm{XYZ}+\sum_{j=1}^L \left(g_j^x\sigma_j^x+g_j^y\sigma_j^y + h_j \sigma_j^z\right),
\end{equation}
implies there is no longer a nontrivial basis in which the Hamiltonian is real and prevents the choice of the \textsc{swap} operator $\swap$ as a pseudo-Hermiticity operator. Since there are no symmetries of the Lindbladian besides $\scT_+$, this case belongs to class AI.

\subsection{Spin injection or removal. Classes BDI and CI}
\label{subsec:examples_InjRem}

We now consider a set of jump operators describing spin injection into the chain (which can occur in the bulk or at the boundaries),
\begin{equation}
L_j=a_j \sigma_j^+,
\end{equation}
where $a_j$ are arbitrary real coefficients. The same considerations apply to the jump operators describing spin removal, $L_j=b_j\sigma_j^-$. In addition, we take the XYZ Hamiltonian of Eq.~(\ref{eq:H_YXZ}), which commutes with all three $\Sigma^{x,y,z}$. Since the jump operators satisfy
\begin{equation}\label{eq:ex_inj_L_Sigma_comm}
\Sigma^x L_j^\dagger \Sigma^x=L_j,
\quad
\Sigma^y L_j^\dagger \Sigma^y=-L_j,
\quad \text{and} \quad
\acomm{L_j}{\Sigma^z}=0,
\end{equation}
we see that the longitudinal parity,
\begin{equation}
\scU^z=\Sigma^z\otimes\Sigma^z,
\end{equation}
is conserved as a Liouvillian weak symmetry, but the transverse parity $\scU^x$ is not. Furthermore, the jump operators satisfy $\acomm{L_j^\dagger}{L_j}=a_j^2 \id_j$, but are neither normal, $\comm{L_j^\dagger}{L_j}=-a_j^2 \sigma^z_j$, nor unitary, $L_j^\dagger L_j=a_j^2(\id_j-\sigma_j^z)/2$. $\scLD'$ can, therefore, only satisfy a $\scQ_-$ symmetry [according to Eqs.~(\ref{eq:sym_scLJ_P}), (\ref{eq:sym_scLJ_Q-}), and (\ref{eq:sym_scLJ_Q+})]. We take 
\begin{equation}
    \scQ_-=\Sigma^x\otimes \Sigma^y
\end{equation}
as the pseudo-Hermiticity superoperator, which satisfies the commutation relation:
\begin{equation}
    \scQ_- \scT_+=(-1)^L\scU^z\, \scT_+ \scQ_-.
\end{equation}
For even $L$, the spin-injection Lindbladian belongs to class BDI in the even parity sector $\scU^z=\scI$ and to class CI in the odd parity sector ($\scU^z=-\scI$). For odd $L$, the result is reversed.

\subsection{Incoherent hopping. Class BDI$^\dagger$ and beyond the tenfold way}
\label{subsec:examples_IncHop}

Next, we consider jump operators describing a two-site XY interaction,
\begin{equation}
\label{eq:jumpops_inchop}
L_{jk}=M_{jk}^x \sigma^x_j\sigma^x_k
+ M_{jk}^y \sigma^y_j\sigma^y_k,
\end{equation}
with arbitrary complex couplings $M_{jk}^\alpha$. In the case $M^x_{jk}=M^y_{jk}$, they describe incoherent hopping. The jump operators satisfy
\begin{equation}
\comm{L_{jk}}{\Sigma^x}=\comm{L_{jk}}{\Sigma^y}=\comm{L_{jk}}{\Sigma^z}=0.
\end{equation}
Choosing the XYZ Hamiltonian in a longitudinal field,
\begin{equation}
H=H_\mathrm{XYZ}+\sum_{j=1}^L h_j^z \sigma^z_j,
\end{equation}
that satisfies
\begin{equation}
\comm{H}{\Sigma^z}=0,
\end{equation}
the longitudinal parity $\Sigma^z$ is a Liouvillian strong symmetry, i.e., the Lindbladian conserves independently left and right longitudinal parity, $\scU^z_\mathrm{L}\scL \scU^z_\mathrm{L}=\scL$ and $\scU^z_\mathrm{R}\scL \scU^z_\mathrm{R}=\scL$, with
\begin{equation}
    \scU^z_\mathrm{L}=\Sigma^z\otimes \id
    \quad \text{and} \quad
    \scU^z_\mathrm{R}=\id\otimes \Sigma^z.
\end{equation}
As discussed in Sec.~\ref{subsec:unitary_syms}, a Lindbladian with a strong symmetry preserves the $\scT_+$ symmetry in steady-state sectors and breaks it in all others. In this case, there is thus a $\scT_+^2=+1$ symmetry in the sectors with even total longitudinal parity, $\scU^z=\scU^z_\mathrm{L}\scU^z_\mathrm{R}=+\scI$, while it is broken for odd total parity, $\scU^z=-\scI$.

Because the jump operators are normal but not unitary [i.e., satisfy Eq.~(\ref{eq:sym_scLJ_Q+}), but not Eq.~(\ref{eq:sym_scLJ_P}) or (\ref{eq:sym_scLJ_Q-})], $\scLD$ only admits a $\scQ_+$ symmetry. Because, additionally, the Hamiltonian and jump operators are symmetric, the pseudo-Hermiticity superoperator is given by $\scQ_+=\swap$ and it commutes with the $\scT_+$ operator as stated in Eq.~(\ref{eq:ex_deph_comm_TQ}). Then, it follows that for even parity, $\scU^z=+\scI$, the Lindbladian belongs to class BDI$^\dagger$, while for odd parity, $\scU^z=-\scI$, it belongs to class AI$^\dagger$, which is outside the tenfold classification of Table~\ref{tab:Lindbladian classes}.

\subsection{Simultaneous dephasing and incoherent hopping. Classes AI$_+$ and AI$_-$}
\label{subsec:examples_Chiral}

We now consider dephasing and incoherent hopping to occur simultaneously and choose jump operators
\begin{align}
\label{eq:jumpops_chiral}
    L_{jk\ell}=\sqrt{\gamma_{jk\ell}}\(\sigma^z_{j}+\eta_{jk\ell}\sigma^x_k\sigma^y_\ell\),
\end{align}
with real $\gamma_{jk\ell}$ and complex $\eta_{jk\ell}$, which satisfy
\begin{equation}
    \acomm{L_{jk\ell}}{\Sigma^x}=\acomm{L_{jk\ell}}{\Sigma^y}=\comm{L_{jk\ell}}{\Sigma^z}=0.
\end{equation}

We take the same Hamiltonian as in Eq.~(\ref{eq:H_deph_noTRS}), which satisfies the commutation relations of Eq.~(\ref{eq:ex_deph_H_Sigma_comm}). This Lindbladian again conserves transverse parity $\scU^x$ as a weak Liouvillian symmetry and has two symmetry sectors.

\begin{figure}[b]
    \centering
    \includegraphics[width=\columnwidth]{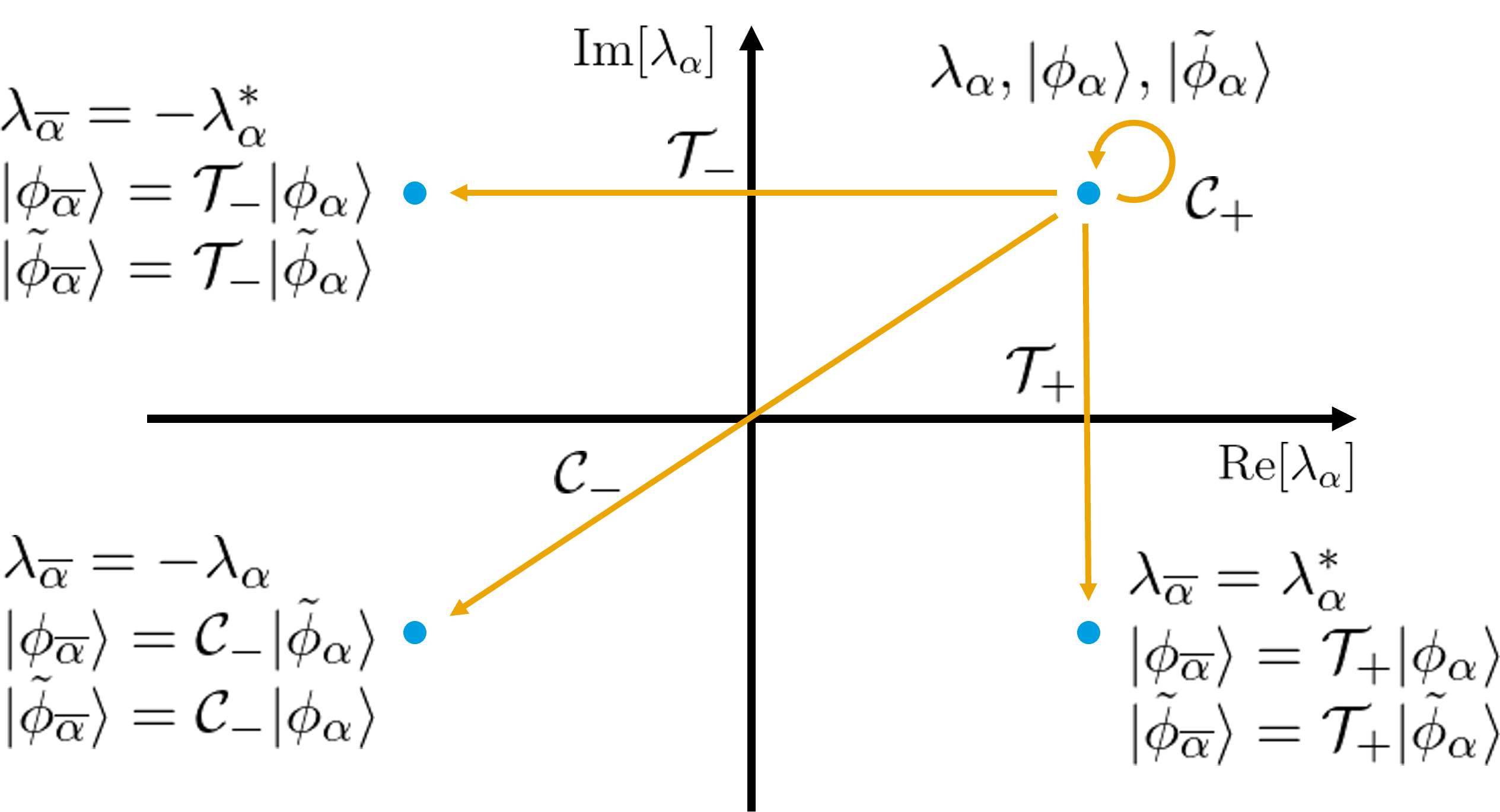}
    \caption{Schematic action of the antiunitary symmetries $\scT_\pm$ and $\scC_\pm$ on the eigenvalues and eigenvectors of $\scL'$. The eigenvalue problem is defined in Eqs.~(\ref{eq:evL}) and (\ref{eq:evLd}) and we depict a representative eigenvalue $\lambda_\alpha$ in the complex plane, together with its image $\lambda_\balpha$ under the symmetries $\scT_\pm$ and $\scC_\pm$. $\scT_+$, $\scT_-$, and $\scC_-$ reflect the spectrum across the real axis, imaginary axis, and origin, respectively, while $\scC_+$ maps an eigenvalue to itself. $\scT_\pm$ map right eigenvectors to right eigenvectors (and left eigenvectors to left eigenvectors), while $\scC_\pm$ map left eigenvectors to right eigenvectors (and vice versa).}
    \label{fig:spectral_consequences}
\end{figure}

When either $j$ equals one of $k$, $\ell$, or $\Re\eta_{jk\ell}=0$, the jump operators are, up to a numerical prefactor, unitary (and, by consequence, normal), and hence, according to Eqs.~(\ref{eq:sym_scLJ_P}), (\ref{eq:sym_scLJ_Q-}), and (\ref{eq:sym_scLJ_Q+}), $\scLD$ admits all of $\scP$ and $\scQ_\pm$ symmetries. However, since the Hamiltonian is not time-reversal symmetric (i.e., is not real in any basis that is trivially related to the representation basis), a $\scQ_-$ symmetry of $\scLH$ requires that both operators $q_a$ and $q_b$ in $\scQ_{ab}=q_a\otimes q_b$ (recall Sec.~\ref{subsec:sym_scLH}) commute with $H$, which would imply $q_a=q_b=\Sigma^x$. This is, however incompatible with a $\scQ_-$ symmetry of $\scLJ$, since it would require that one of $q_a$, $q_b$ commutes with the jump operators and the other one anticommutes (recall Sec.~\ref{subsec:sym_scLJ}). Noting that the jump operators are non-Hermitian, we can also exclude a $\scQ_+$ symmetry. We thus conclude that the Lindbladian only possesses $\scP$ symmetry, which is implemented by the unitary operator:
\begin{equation}
    \scP=\Sigma^z\otimes\Sigma^y.
\end{equation}
Because of the commutation relation with the $\scT_+$ symmetry, 
\begin{equation}
    \scP \scT_+ = (-1)^L \scU^x \, \scT_+ \scP,
\end{equation}
we find that the Liouvillian belongs to class AI$_+$ if $L$ and the parity $\scU^x$ are both even or both odd, and to class AI$_-$ if one of $L$ or $\scU^x$ is even and the other odd.

\begin{figure*}[t]
    \centering
    \includegraphics[width=\textwidth]{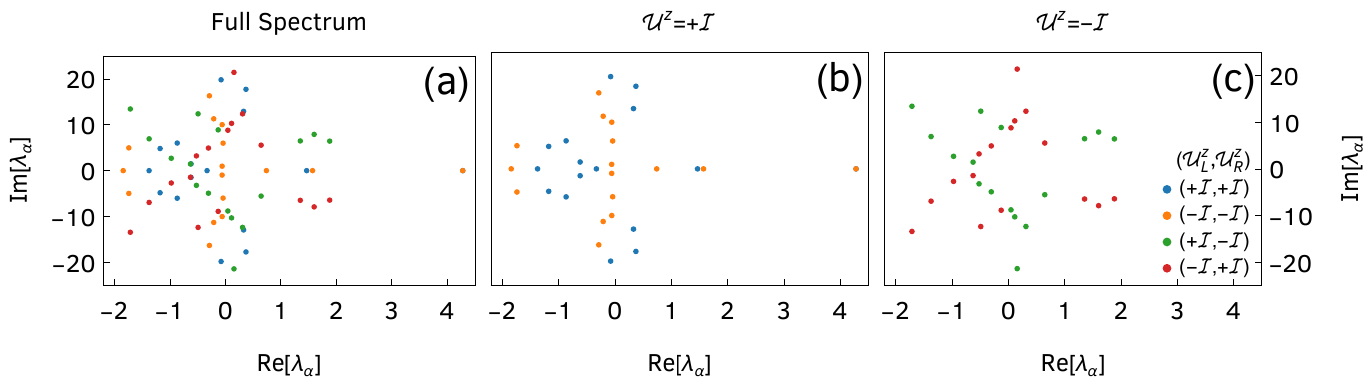}
    \caption{Spectrum of $\scL'$ in the complex plane for the incoherent hopping chain of length $L=3$ discussed in Sec.~\ref{subsec:examples_IncHop}. The remaining parameters are given in Appendix~\ref{app:numerics}. The Lindbladian has four strong-symmetry sectors labeled by the pair $(\scU^z_\mathrm{L},\scU^z_\mathrm{R})$ and represented in different colors. For visual clarity, we present the full spectrum in (a), the two sectors with even parity---for which $\scT_+$ is unbroken, contain a steady-state each, and eigenvalues come in complex-conjugated pairs in each---in (b), and the two sectors with odd parity---for which $\scT_+$ is broken---in (c).}
    \label{fig:Strong_Symm}
\end{figure*}

\section{Random-matrix correlations and universality}
\label{sec:rmt}

Having established the tenfold classification of irreducible Lindbladians and presented physical examples of all classes, we now look for signatures of random-matrix universality in each of those classes. As we have seen above, a Lindbladian class can be labeled by its antiunitary symmetries $\scT_-$ and $\scC_\pm$ or, equivalently, the closely related unitary involutions $\scP$ and $\scQ_\pm$, see Eq.~(\ref{eq:antiunitary_from_unitary}). In this section, we will use $\scT_-$ and $\scC_\pm$.

\subsection{Spectral consequences of antiunitary symmetries}
\label{subsec:spectrum}

We start by reviewing the constraints the antiunitary symmetries $\scT_\pm$ and $\scC_\pm$ impose on the eigenvalues and eigenvectors of the Lindbladians~\cite{hamazaki2019}, which are schematically summarized in Fig.~\ref{fig:spectral_consequences}.

We argued in Sec.~\ref{sec:classification} that $\scC_-$ and $\scT_-$ symmetries reflect the spectrum of $\scL$ across the origin or imaginary axis. Let us make this statement more precise. We denote the eigenvalues of the vectorized shifted Lindbladian by $\lambda_\alpha$ and the, in general distinct, right and left eigenvectors by $\ket{\phi_\alpha}$ and $\ket{\tphi_\alpha}$, respectively, i.e., 
\begin{align}
    \label{eq:evL}
    &\scL' \ket{\phi_\alpha}=\lambda_\alpha \ket{\phi_\alpha},
    \\
    \label{eq:evLd}
    &\scL'^\dagger \ket{\tphi_\alpha}=\lambda_\alpha^* \ket{\tphi_\alpha}.
\end{align}

Let us first consider the presence of a $\scT_+$ symmetry. Applying $\scT_+$ to Eq.~(\ref{eq:evL}) and using Eq.~(\ref{eq:nHsym_Tp}), we obtain
\begin{equation}
    \scL' \left(\scT_+\ket{\phi_\alpha}\right)=
    \lambda_\alpha^* \left(\scT_+\ket{\phi_\alpha}\right),
\end{equation}
that is, $\scT_+\ket{\phi_\alpha}$ is also a right eigenvector of $\scL'$ with complex-conjugated eigenvalue. When $\scT_+$ is unbroken, the spectrum of $\scL'$ is symmetric about the real axis. This is illustrated in Fig.~\ref{fig:Strong_Symm}, where we show the spectrum of $\scL'$ in the complex plane for the incoherent hopping example of Sec.~\ref{sec:examples}. Recall that this example has four strong-symmetry sectors, labeled by the pair of longitudinal parities $(\scU^z_\mathrm{L},\scU^z_\mathrm{R})$. As is clearly visible, and in agreement with our predictions, the full spectrum; see Fig.~\ref{fig:Strong_Symm}(a), and the spectra of the two sectors with total parity $\scU^z=\scU^z_\mathrm{L}\scU^z_\mathrm{R}=+\scI$, see Fig.~\ref{fig:Strong_Symm}(b), are symmetric about the real axis since $\scT_+$ is unbroken; while $\scT_+$ connects the two sectors with $\scU^z=-\scI$ and does not act inside of each (i.e., is broken), leading to each pair of complex-conjugated eigenvalues to be split between the two sectors, see Fig.~\ref{fig:Strong_Symm}(c).

Proceeding similarly, we find that if there is a $\scT_-$ symmetry, $\scT_-\ket{\phi_\alpha}$ is a right eigenvector with eigenvalue $-\lambda_\alpha^*$; i.e., the spectrum is symmetric about the imaginary axis. We conclude that the presence of $\scT_-$ leads to a dihedral symmetry of the Lindbladian spectrum, a phenomenon first pointed out in Ref.~\cite{prosen2012PRL}. 

If there is an antiunitary symmetry $\scC_\pm$, then the operator implementing it connects left and right eigenvectors. Indeed, for $\scC_+$ we have
\begin{equation}
    \scL'^\dagger \left(\scC_+\ket{\phi_\alpha}\right)=
    \lambda_\alpha^* \left(\scC_+\ket{\phi_\alpha}\right);
\end{equation}
that is, $\scC_+\ket{\phi_\alpha}$ is a left eigenvector of $\scL'$ with the same eigenvalue as $\ket{\phi_\alpha}$. Accordingly, this symmetry does not affect the global shape of the spectrum. Furthermore, if $\scC_+^2=-1$, then each eigenvalue is doubly degenerate, a phenomenon dubbed non-Hermitian Kramers degeneracy. 

Finally, if there is a $\scC_-$ symmetry, $\scC_-\ket{\phi_\alpha}$ is a left eigenvector of $\scL'$ with eigenvalue $-\lambda_\alpha$; i.e., $\scC_-$ reflects the spectrum across the origin. Therefore, the presence of this symmetry also implies the dihedral symmetry of the spectrum.

\begin{figure}[t]
    \centering
    \includegraphics[width=\columnwidth]{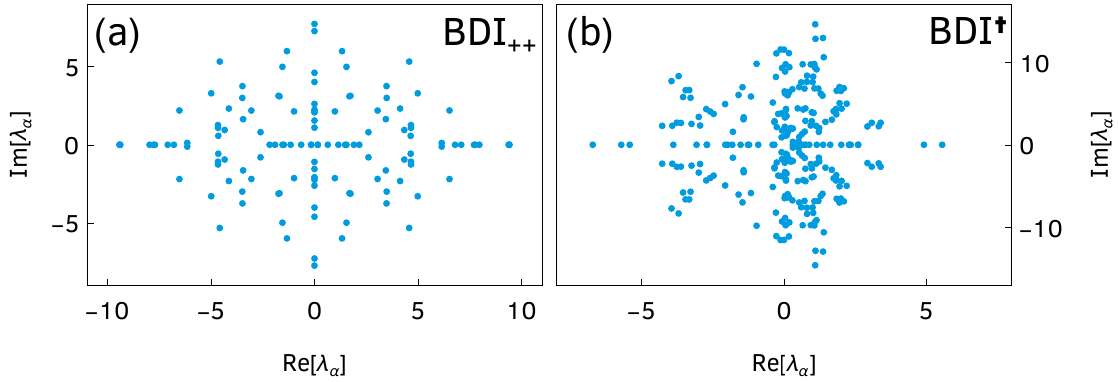}
    \caption{Spectrum of $\scL'$ in the complex plane for (a) the dephasing spin chain with Hamiltonian $H_\mathrm{XYZ}+H_\mathrm{X}$ (Sec.~\ref{subsec:examples_Deph}) and (b) the incoherent hopping chain (Sec.~\ref{subsec:examples_IncHop}). In both cases, we consider $L=4$ and the remaining parameters are given in Appendix~\ref{app:numerics}. The dephasing spin chain belongs to class BDI$_{++}$ and the spectrum of the shifted Lindbladian has dihedral symmetry, while the incoherent hopping chain belongs to class BDI$^\dagger$ and its spectrum does not display dihedral symmetry.}
    \label{fig:spectra}
\end{figure}

From the data in Table~\ref{tab:Lindbladian classes} and the preceding discussion, we conclude there are eight classes with dihedral symmetry and two without (AI and BDI$^\dagger$). We illustrate this in Fig.~\ref{fig:spectra}, where we show the spectrum of $\scL'$ in the complex plane for two examples of Sec.~\ref{sec:examples}: the dephasing spin chain belonging to class BDI$_{++}$, which presents dihedral symmetry, and the incoherent hopping chain in class BDI$^\dagger$, which does not.

\subsection{Complex spacing ratios}

We now move to the random matrix signatures of the different antiunitary symmetries. First, we consider (bulk) local level statistics, which are sensitive to the value of $\scC_+^2$ (we denote the absence of the symmetry as $\scC_+^2=0$). Local level statistics are most conveniently captured by the distribution of complex spacing ratios (CSRs)~\cite{sa2019csr}. (The alternatives, the bare complex spacing distribution~\cite{akemann2019} and the dissipative spectral form factor~\cite{li2021PRL} require a cumbersome unfolding procedure~\cite{garcia2022ARXIV}.) CSRs have become a popular measure of dissipative quantum chaos, ranging from studies of random Lindbladians~\cite{wang2020,tarnowski2021,sa2022,costa2022} to nonunitary quantum circuits~\cite{sa2021}, non-Hermitian Anderson transitions~\cite{huang2020,luo2021}, and two-color QCD~\cite{kanazawa2021}, among others.
We define the CSR as~\cite{sa2019csr}
\begin{equation}
    z_\alpha=\frac{
    \lambda_\alpha^{\mathrm{NN}}-\lambda_\alpha}{
    \lambda_\alpha^{\mathrm{NNN}}-\lambda_\alpha
    },
\end{equation}
where $\lambda_\alpha^{\mathrm{NN}}$ and $\lambda_\alpha^{\mathrm{NNN}}$ are the nearest and next-to-nearest neighbors of $\lambda_\alpha$ in the complex plane. By definition, $z_\alpha$ is constrained to the unit disk. Since they are defined in terms of the two nearest eigenvalues, CSRs only measure correlations up to a few level spacings. As a consequence, they can only be sensitive to the symmetry $\scC_+$. Indeed, the other three antiunitary symmetries correlate eigenvalues that are, in a many-body system, exponentially many level spacings apart (reflected across the real or imaginary axis, or the origin).

\begin{figure}[t]
    \centering
    \includegraphics[width=\columnwidth]{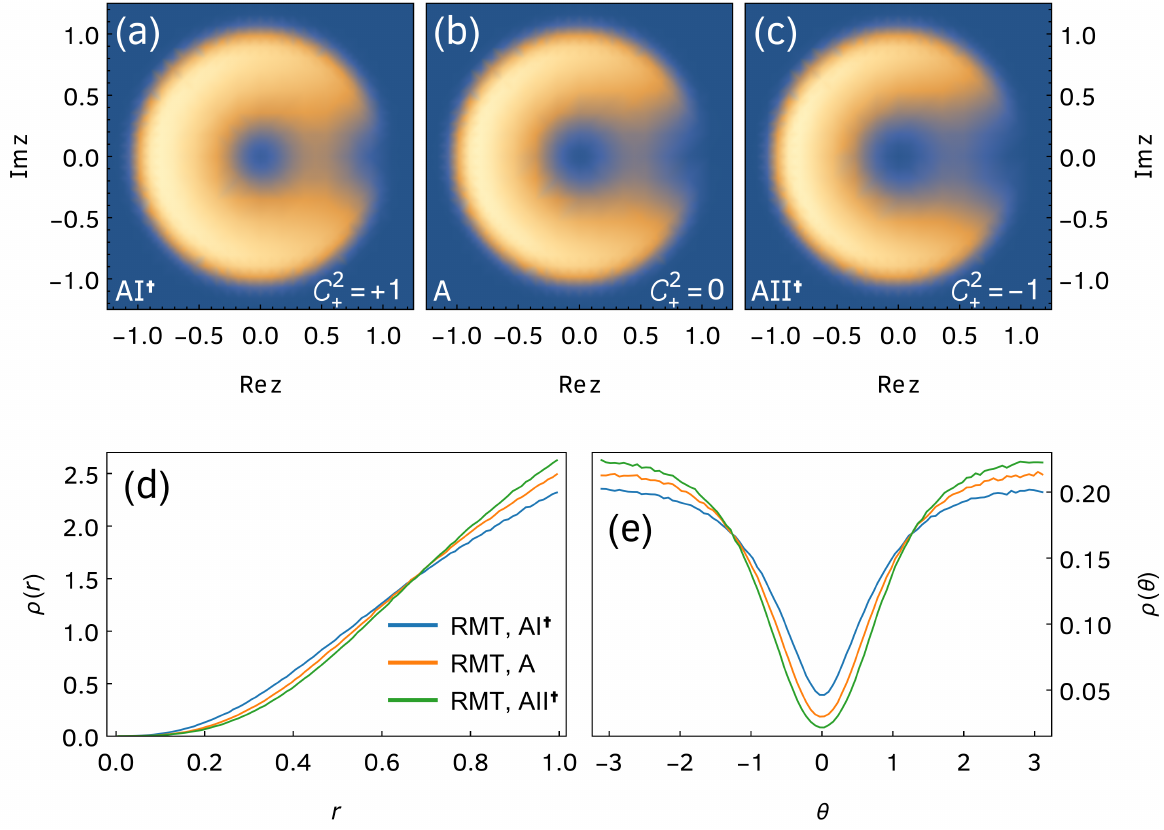}
    \caption{Complex spacing ratio distribution of random matrices in the three bulk classes A, AI$^\dagger$, and AII$^\dagger$. The distribution of $z$ in the complex plane (a)--(c) has a characteristic donut-like shape. The hole at the origin and the low probability at small angles $\theta=0$ are a sign of level repulsion and increase from class AI$^\dagger$ to A to AII$^\dagger$. We obtain these distributions numerically from exact diagonalization of an ensemble of $2^{15}\times2^{15}$ random matrices with $2^8$ realizations. We note that good analytical approximations exist for class A~\cite{sa2020,dusa2022}, but not for classes AI$^\dagger$ and AII$^\dagger$. A more quantitative comparison can be done by studying the marginal (d) radial and (e) angular distributions. These distributions are compared against the physical spin-chain results in Fig.~\ref{fig:CSR_SpinChains}.}
    \label{fig:CSR_RMT}
\end{figure}

For random matrices, the CSR distribution acquires a characteristic donut-like shape, with the details of the distribution only dependent on the value of $\scC_+^2$; see Figs.~\ref{fig:CSR_RMT}(a)--(c). The three types of level repulsion are usually denoted as A ($\scC_+^2=0$), AI$^\dagger$ ($\scC_+^2=+1$), and AII$^\dagger$ ($\scC_+^2=-1$)~\cite{hamazaki2019}. Level repulsion in class AII$^\dagger$ does not occur in Lindbladian symmetry classes.

\begin{figure*}[t]
    \centering
    \includegraphics[width=\textwidth]{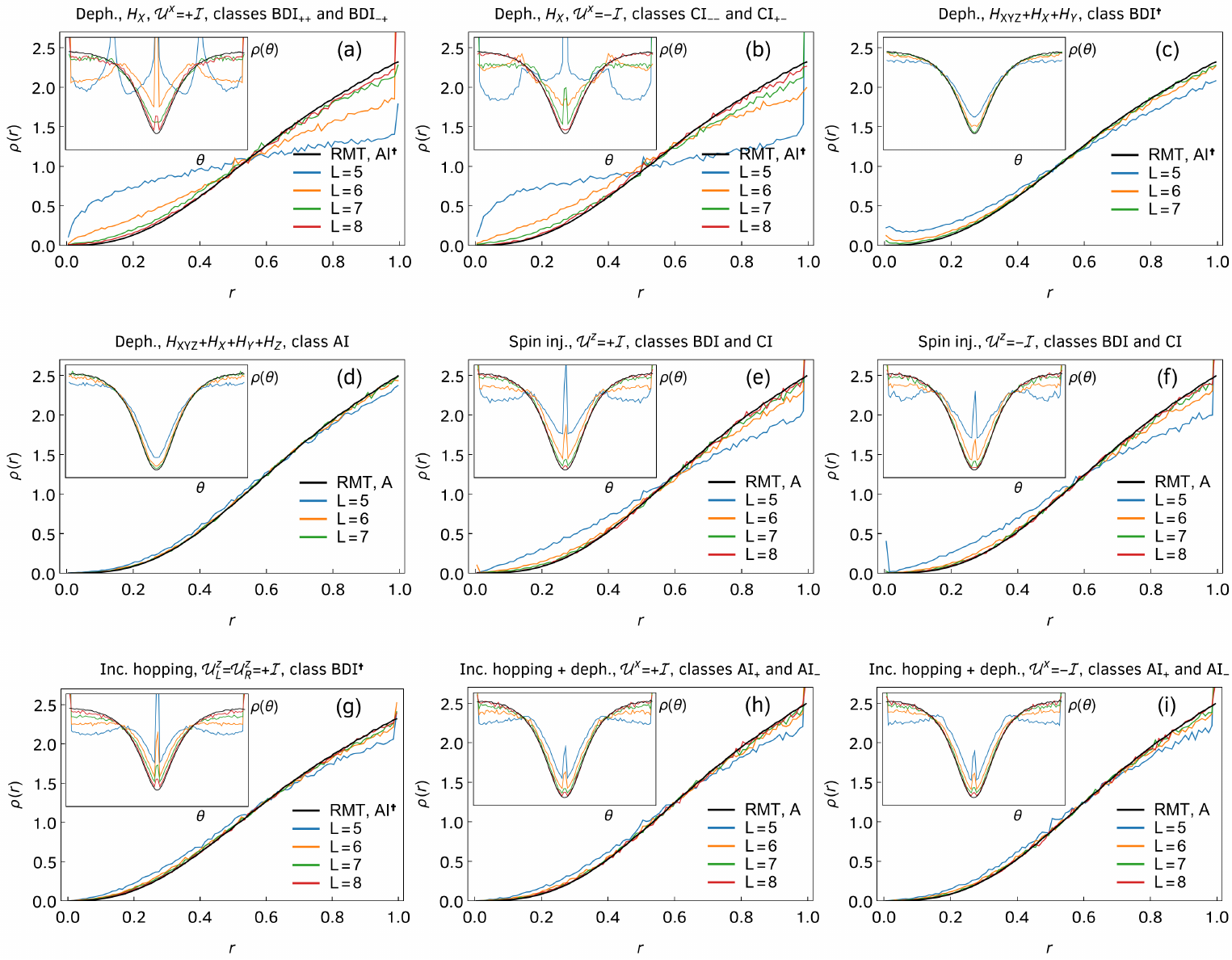}
    \caption{Complex spacing ratio distribution of all the spin-chain examples discussed in Sec.~\ref{sec:examples}. In each panel, we show the marginal radial distribution for different chain lengths (colored lines) and compare it with the random matrix prediction of Fig.~\ref{fig:CSR_RMT} (black line). In the insets, we show the marginal angular distribution. In all cases, we observe excellent agreement with the universal RMT result as $L$ increases, while there are also very strong finite-size effects that difficult a comparison for $L=5$ and $6$.}
    \label{fig:CSR_SpinChains}
\end{figure*}

To identify random-matrix universality in the examples of Sec.~\ref{sec:examples}, we randomly sample disordered spin chains and compute the CSR distribution. Details on the numerical simulations, including the values of the parameters for each example, are given in Appendix~\ref{app:numerics}. To make a more quantitative comparison with the random matrix theory (RMT) results, it is convenient to consider the marginal radial, $\rho(r)$, and angular, $\rho(\theta)$, distributions of the CSR expressed as $z_\alpha=r_\alpha \exp\{i \theta_\alpha\}$. They are shown in Figs.~\ref{fig:CSR_RMT}(d) and (e) for the three bulk RMT ensembles. In Fig.~\ref{fig:CSR_SpinChains}, we compare them with the marginal distributions for all the physical spin-chain examples discussed in Sec.~\ref{sec:examples}, finding excellent agreement with RMT predictions when the length $L$ of the chain becomes large. Our results illustrate RMT universality in the full tenfold classification of Lindbladians with unbroken $\scT_+$ symmetry. 

Through the use of bulk CSR, we can only resolve the value of $\scC_+^2$ (which is manifest in some panels of Fig.~\ref{fig:CSR_SpinChains}, where we group together results for different symmetry classes that share the same level repulsion). We now discuss numerical signatures that can also distinguish the values of $\scC_-^2$ and $\scT_-^2$.

\subsection{Statistics of real and imaginary eigenvalues and eigenvalues close to the origin}

\begin{figure*}[t]
    \centering
    \includegraphics[width=\textwidth]{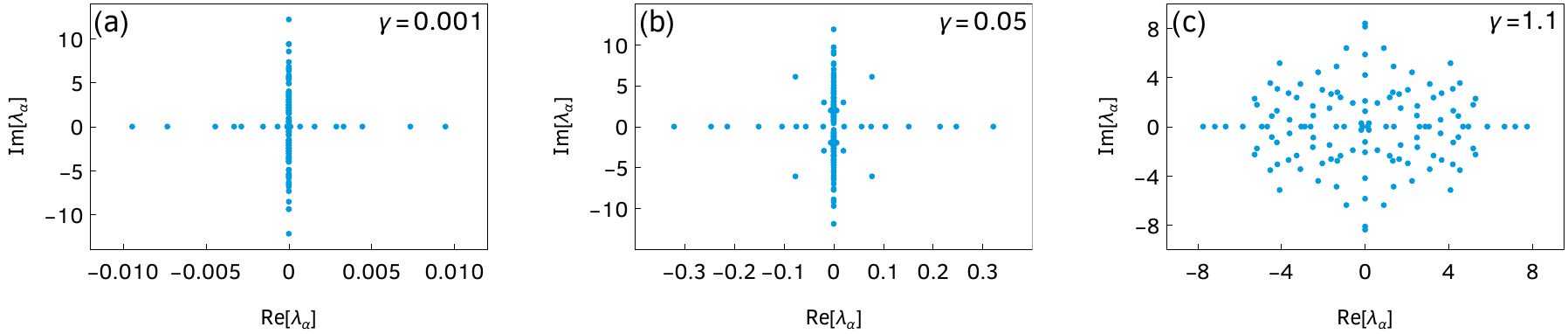}
    \caption{Spectrum of $\scL'$ in the complex plane for the dephasing spin chain with Hamiltonian $H_\mathrm{XYZ}+H_\mathrm{X}$ (Sec.~\ref{subsec:examples_Deph}) for $L=4$ and different dephasing strengths $\gamma$. The remaining parameters are given in Appendix~\ref{app:numerics}. For small $\gamma$ (a), PT symmetry is unbroken and the whole spectrum lives on the real and imaginary axes. For large enough $\gamma$ (b), a series of exceptional points occurs, with a pair of eigenvalues on one of the symmetry axes colliding and shooting off into the complex plane, spontaneously breaking PT symmetry. For very large $\gamma$ (c), most of the spectrum lives in the complex plane. This phenomenon of spontaneous PT symmetry breaking renders the number of real and purely imaginary eigenvalues, and their statistics, nonuniversal, and we do not use them to characterize the symmetries and correlations Lindbladians.}
    \label{fig:PT_symmetry_breaking}
\end{figure*}

In Hermitian systems, particle-hole and chiral symmetries manifest themselves in the eigenvalues near the origin, because these symmetries reflect the spectrum across it. Not only are universal local correlations in this region distinct from the bulk, even the spectral density is universal in a certain microscopic limit and determined only by the symmetry~\cite{verbaarschot1993,verbaarschot1994,akemann1997}. The same reasoning leads to the speculation that in order to obtain local information on $\scC_-$ and $\scT_\pm$ symmetries, we need to restrict our attention to the vicinity of the axis of symmetry of the spectrum. The effects of these symmetries on the correlations near the origin have been addressed in the Ginibre~\cite{akemann2009JPhysA,akemann2009PRE} and non-Ginibre classes~\cite{garcia2022PRX}, while universal statistics of real eigenvalues were studied for the real Ginibre~\cite{lehmann1991PRL,kanzieper2005PRL,forrester2007PRL} and non-Ginibre classes~\cite{xiao2022}.

First, we point out that the statistics of the eigenvalues closest to the origin, employed in Ref.~\cite{garcia2022PRX} as a signature of different symmetry classes, are affected by the existence of exact zero modes of $\scL'$ in some of our families of spin-1/2 chains. The existence of these zero modes for all realizations of disorder induces additional level repulsion from the origin, altering the distribution of nonzero eigenvalues closest to it. As an example of this phenomenon, we mention the chain with spin injection, belonging to class BDI, which supports two exact zero modes. In principle, if the number of zero modes does not scale with system size, one could study the distribution of the eigenvalues closest to the origin for each number of zero modes.

More critically, we also find that for our examples in disordered spin chains, the statistics of eigenvalues on or near the axes of symmetry are nonuniversal because of the spontaneous breaking of PT symmetry~\cite{prosen2012PRL}. In Ref.~\cite{xiao2022} it was found that, in the ergodic regime (in which RMT behavior is expected), the number of real eigenvalues in the spectra of some physical non-Hermitian Hamiltonians is universal and equal to the random matrix value $~\propto\sqrt{D}$, with $D$ the Hilbert space dimension. In contrast, for Lindbladians with dihedral symmetry, the fraction of real and imaginary eigenvalues and its statistics depend on the relative strength $g$ of the non-Hamiltonian part of the Lindbladian ($g$ can be, for instance, the dephasing strength $\gamma$ or the spin-injection rate $a$). For $g<g_{\mathrm{PT}}$, with finite size-dependent critical $g_\mathrm{PT}$, all eigenvalues of the shifted Lindbladian $\mathcal L'$ reside on the cross formed by the real and imaginary axes (PT-unbroken phase)~\cite{prosen2012PRL}. At $g=g_\mathrm{PT}$ (the first exceptional point), a pair of eigenvalues on the cross collides and shoots off into the complex plane, spontaneously breaking PT symmetry, and, consequently, reducing the number of real or imaginary eigenvalues. As $g$ increases further, more collisions of eigenvalues occur. The change in the number of real eigenvalues for the dephasing spin chain in class BDI$_{++}$ as a function of $\gamma$ is illustrated in Fig.~\ref{fig:PT_symmetry_breaking}. 
Concomitantly with the nonuniversality of the number of eigenvalues on the axes of symmetry, we also find their statistics to be nonuniversal and depend sensitively on the coupling $g$. As a consequence, we are not able to employ the statistics of, say, purely imaginary eigenvalues as a diagnostic of the symmetries and RMT universality in Lindbladian classes.
We expect that for $g\gg g_\mathrm{PT}$, i.e., deep in the symmetry-broken phase, the number of real and imaginary eigenvalues becomes universal and their statistics obey RMT. In particular, in the thermodynamic limit, we expect $g_\mathrm{PT}\to0$~\cite{prosen2012PRL}, and, hence RMT statistics for all nonzero dissipation. However, since we have only access to relatively small system sizes and the disorder changes the precise value of $g_\mathrm{PT}$ from realization to realization, we do not pursue this question further in this work, and instead turn to an alternative signature of the different symmetries that works at any coupling $g$.

\subsection{Eigenvector overlaps}

Having ruled out the prospect of inferring antiunitary symmetries of Lindbladians through local spectral information near the symmetry axis, we turn to the possibility of using nonlocal bulk information. To that end, we consider the Chalker-Mehlig eigenvector overlap matrix~\cite{mehlig1998PRL,mehlig2000JMP}:
\begin{equation}
 O_{\alpha\beta}=\braket{\tphi_\alpha}{\tphi_\beta}\braket{\phi_\beta}{\phi_\alpha}.
\end{equation}
Note that this definition applies only in the case of classes without non-Hermitian Kramers degeneracy. If each eigenvalue is doubly degenerate and the eigenspace dimension is two, each element $O_{\alpha\beta}$ becomes itself a $4\times4$ matrix, and additional care has to be exerted. Since classes with Kramers degeneracy do not occur in the Lindbladian classification, we defer such considerations to future work and do not consider this case further in this paper.

\begin{figure*}[t]
    \centering
    \includegraphics[width=0.8\textwidth]{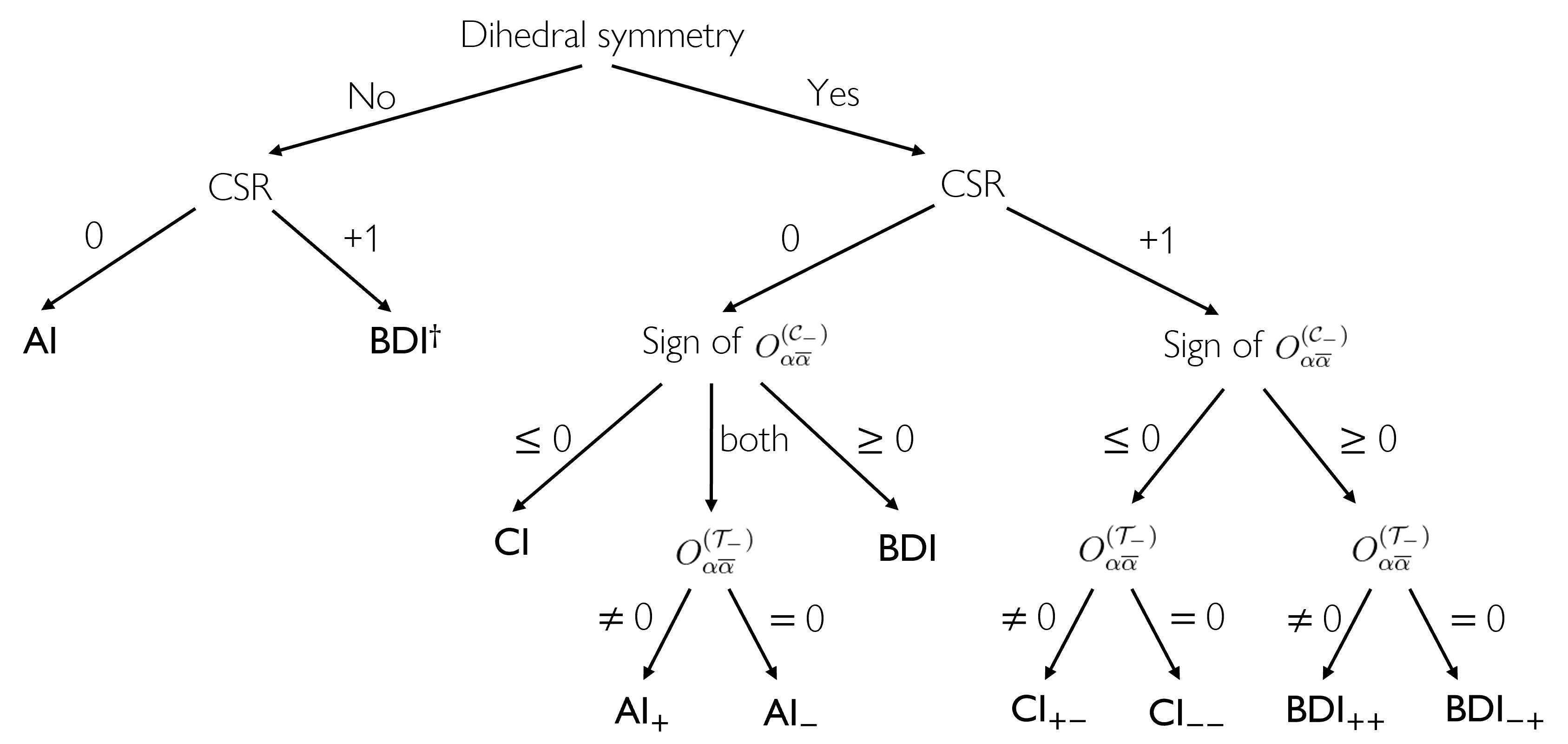}
    \caption{Decision tree illustrating the possibility of distinguishing the full Lindbladian tenfold classification by jointly employing the dihedral symmetry of the spectrum, the complex spacing ratio distribution (denoted as $0$ for bulk level repulsion of class A and $+1$ for class AI$^\dagger$), the sign of the off-diagonal eigenvector overlap $O^{(\scC_-)}_{\alpha\balpha}$, and whether or not the overlaps $O^{(\scT_-)}_{\alpha\balpha}$ are identically zero.}
    \label{fig:decision_tree_classes}
\end{figure*}

The overlap matrix, in particular the distribution of its entries and the first moments, have been intensely investigated for the Ginibre ensembles~\cite{mehlig1998PRL,mehlig2000JMP,janik1999PRE,fyodorov2018,bourgade2020,akemann2020Acta}. The diagonal overlaps $O_{\alpha\alpha}$ are sensitive to $\scC_+^2$, a claim we have confirmed numerically. However, since local eigenvalue statistics---measured, for instance, by CSR---are already sensitive to this symmetry, we do not employ it in this paper.

Instead, we propose that the off-diagonal overlaps $O_{\alpha\balpha}$, where $\{\ket{\phi_\alpha},\ket{\tphi_\alpha}\}$ and $\{\ket{\phi_\balpha},\ket{\tphi_\balpha}\}$ are connected by an antiunitary symmetry $\scT_\pm$ or $\scC_-$, are sensitive to the value of the square of that symmetry. More concretely, we begin by making the following empirical observations for random matrices.
\begin{enumerate}
    \item If $\ket{\tphi_\balpha}\propto \scC_-\ket{\phi_\alpha}$, the overlaps $O_{\alpha\balpha}$ (denoted $O_{\alpha\balpha}^{(\scC_-)}$ for clarity) are all non-negative if $\scC_-^2=+1$, and all nonpositive if $\scC_-^2=-1$. If $\scC_-^2=0$ (i.e., if the symmetry is absent and the eigenvectors are independent), the overlaps are still real for spectra with dihedral symmetry, and the fraction of positive and negative matrix elements is $1/2$ each. For classes with no $\scC_-$ symmetry and no dihedral symmetry, the overlaps are complex.
    \item If $\ket{\phi_\balpha}\propto \scT_-\ket{\phi_\alpha}$ and $\scT_-^2=-1$, the overlaps $O_{\alpha\balpha}$ (denoted $O_{\alpha\balpha}^{(\scT_-)}$) all vanish identically. If $\scT_-^2=+1$ or $0$ they assume arbitrary complex values.
\end{enumerate}

{
\setlength{\tabcolsep}{8pt}
\begin{table*}[t]
\caption{Signatures of $\scC_-$ and $\scT_-$ symmetries in eigenvector overlaps for all the spin-chain examples discussed in Sec.~\ref{sec:examples}. Each set of examples (dephasing, spin injection or removal, incoherent hopping, and simultaneous incoherent hopping and dephasing) realizes different classes depending on the choice of Hamiltonian, conserved parity sectors, and the parity of the chain length $L$. The examples are listed in the same order as discussed in Sec.~\ref{sec:examples}. For each, we give the corresponding symmetry class, the values of the square of the two antiunitary symmetries $\scC_-$ and $\scT_-$, whether the sign of the off-diagonal overlap $O_{\alpha\balpha}^{(\scC_-)}$ is mostly non-negative or nonpositive, together with the fraction of times this happens, and whether the off-diagonal overlap $O_{\alpha\balpha}^{(\scT_-)}$ is mostly zero or nonzero, together with the fraction of times this happens. We see that the criteria we put forward for the values of $\scC_-^2$ and $\scT_-^2$ are always satisfied at least $99.9\%$ of the time and all the predictions of Sec.~\ref{sec:examples} are verified.}
\label{tab:overlap_results}
\begin{tabular}{@{}llrrll@{}}
\toprule
Example                                                        & Class         & $\scC_-^2$ & $\scT_-^2$ & $\mathrm{Re}\,O_{\alpha\balpha}^{(\scC_-)}$ & $O_{\alpha\balpha}^{(\scT_-)}$ \\ \midrule
Deph., $H_\mathrm{X}$, $\scU^x=+\scI$, $L$ even                 & BDI$_{++}$    & $+1$       & $+1$       & $\geq0$, $100\%$               & $\neq0$, $99.98\%$             \\
Deph., $H_\mathrm{X}$, $\scU^x=+\scI$, $L$ odd                  & BDI$_{-+}$    & $+1$       & $-1$       & $\geq0$, $100\%$               & $=0$, $99.98\%$                \\
Deph., $H_\mathrm{X}$, $\scU^x=-\scI$, $L$ even                 & CI$_{--}$     & $-1$       & $-1$       & $\leq0$, $100\%$               & $=0$, $99.92\%$                \\
Deph., $H_\mathrm{X}$, $\scU^x=-\scI$, $L$ odd                  & CI$_{+-}$     & $-1$       & $+1$       & $\leq0$, $100\%$               & $\neq0$, $99.99\%$             \\
Deph., $H_\mathrm{XYZ}+H_\mathrm{X}$, $\scU^x=+\scI$            & BDI$_{++}$    & $+1$       & $+1$       & $\geq0$, $100\%$               & $\neq0$, $100\%$               \\
Deph., $H_\mathrm{XYZ}+H_\mathrm{X}$, $\scU^x=-\scI$            & CI$_{+-}$     & $-1$       & $+1$       & $\leq0$, $100\%$               & $\neq0$, $100\%$               \\
Deph., $H_\mathrm{XYZ}+H_\mathrm{X}+H_\mathrm{Y}$              & BDI$^\dagger$ & $0$        & $0$        & $\leq0$, $50.12\%$             & $\neq0$, $100\%$               \\
Deph., $H_\mathrm{XYZ}+H_\mathrm{X}+H_\mathrm{Y}+H_\mathrm{Z}$ & AI            & $0$        & $0$        & $\leq0$, $50.07\%$             & $\neq0$, $99.999\%$            \\
Spin inj., $\scU^z=+\scI$, $L$ even                            & BDI           & $+1$       & $0$        & $\geq0$, $100\%$               & $\neq0$, $100\%$               \\
Spin inj., $\scU^z=+\scI$, $L$ odd                             & CI            & $-1$       & $0$        & $\leq0$, $100\%$               & $\neq0$, $100\%$               \\
Spin inj., $\scU^z=-\scI$, $L$ even                            & CI            & $-1$       & $0$        & $\leq0$, $100\%$               & $\neq0$, $100\%$               \\
Spin inj., $\scU^z=-\scI$, $L$ odd                             & BDI           & $+1$       & $0$        & $\geq0$, $100\%$               & $\neq0$, $100\%$               \\
Inc. hopping, $\scU^z_\mathrm{L}=\scU^z_\mathrm{R}=+\scI$      & BDI$^\dagger$ & $0$        & $0$        & $\leq0$, $51.01\%$             & $\neq0$, $100\%$             \\
Inc. hopping + deph., $\scU^x=+\scI$, $L$ even                 & AI$_+$        & $0$        & $+1$       & $\geq0$, $50.50\%$             & $\neq0$, $100\%$               \\
Inc. hopping + deph., $\scU^x=+\scI$, $L$ odd                  & AI$_-$        & $0$        & $-1$       & $\geq0$, $50.44\%$             & $=0$, $99.98\%$                \\
Inc. hopping + deph., $\scU^x=-\scI$, $L$ even                 & AI$_-$        & $0$        & $-1$       & $\geq0$, $50.06\%$             & $=0$, $99.94\%$                \\
Inc. hopping + deph., $\scU^x=-\scI$, $L$ odd                  & AI$_+$        & $0$        & $+1$       & $\geq0$, $51.20\%$             & $\neq0$, $100\%$               \\ \bottomrule
\end{tabular}
\end{table*}
}

In turn, these two statements can be proven in general by a variation of the proof of Kramers degeneracy. 
Let us denote by $\scA$ one of the four antiunitary operators $\scT_\pm$ or $\scC_\pm$. Then, for any two vectors $\psi$ and $\phi$, we have
\begin{equation}
\label{eq:antiunitary_braket}
\begin{split}
    \braket{\psi}{\scA\phi}
    =\braket{\scA \psi}{\scA^2 \phi}^*
    =\scA^2 \braket{\scA \psi}{\phi}^*
    =\scA^2 \braket{\phi}{\scA \psi},
\end{split}
\end{equation}
where we use the antiunitarity of $\scA$ and the fact that $\scA^2$ is either $\pm 1$.
In order to prove assertion 1, following Sec.~\ref{subsec:spectrum}, we note that 
\begin{equation}
    \ket{\tphi_\balpha}=\scC_-\ket{\phi_\alpha}\,,\quad
    \ket{\phi_\balpha}=\scC_-\ket{\tphi_\alpha}\,,
\end{equation}
where, without loss of generality, we set a possible proportionality constant to one. 
Then, using Eq.~(\ref{eq:antiunitary_braket}), the overlap matrix reads
\begin{equation}
\begin{split}
    O^{(\scC_-)}_{\alpha\balpha}
    &=\bra{\tphi_\alpha}\scC_-\ket{\phi_\alpha}
    \bra{\tphi_\alpha}\scC_-^\dagger\ket{\phi_\alpha}
    \\
    &=\scC_-^2 \bra{\phi_\alpha}\scC_-\ket{\tphi_\alpha}
    \bra{\tphi_\alpha}\scC_-^\dagger\ket{\phi_\alpha}
    \\
    &=\scC_-^2 \abs{\bra{\phi_\alpha}\scC_-\ket{\tphi_\alpha}}^2,
\end{split}
\end{equation}
and we conclude that the overlap matrix element $O^{(\scC_-)}_{\alpha\balpha}$ has the same sign as $\scC_-^2$, proving assertion 1. In order to prove assertion 2, we note instead the relation between the two right eigenvectors:
\begin{equation}
    \ket{\phi_\balpha}=\scT_-\ket{\phi_\alpha}.
\end{equation}
Using Eq.~(\ref{eq:antiunitary_braket}), it immediately follows that
\begin{equation}
    \bra{\phi_\alpha}\scT_- \ket{\phi_\alpha}
    =\scT_-^2\bra{\phi_\alpha}\scT_- \ket{\phi_\alpha}.
\end{equation}
If $\scT_-^2=-1$, this matrix element and, consequently, the overlap 
\begin{equation}
    O^{(\scT_-)}_{\alpha\balpha}=\bra{\tphi_\alpha}\scT_-\ket{\tphi_\alpha}\bra{\phi_\alpha}\scT_-^\dagger\ket{\phi_\alpha}
\end{equation}
vanish identically, proving assertion 2.

Very importantly, explicit knowledge of the operator $\scC_\pm$ or $\scT_\pm$ is not required to compute the respective eigenvector overlaps. To construct the overlap matrix, the eigenvalues are ordered by increasing real part and, for each pair of complex conjugated eigenvalues, by increasing imaginary part. With this ordering, the overlaps $O^{(\scC_+)}_{\alpha\balpha}$ lie on the main diagonal $O_{\alpha\alpha}$ of the matrix $O_{\alpha\beta}$; the overlaps $O^{(\scC_-)}_{\alpha\balpha}$ are the antidiagonal elements $O_{\alpha,D-\alpha+1}$, where $D$ is the sector dimension; the overlaps $O^{(\scT_+)}_{\alpha\balpha}$ are the elements $O_{2\alpha-1,2\alpha}$; and, finally, the overlaps $O^{(\scT_-)}_{\alpha\balpha}$ are the elements $O_{2\alpha-1,D-2\alpha+1}$ and $O_{2\alpha,D-2\alpha+2}$.

The overlaps $O^{(\scC_-)}_{\alpha\balpha}$ and $O^{(\scT_-)}_{\alpha\balpha}$, together with the CSR distribution, are enough to distinguish the ten Lindbladian classes with unbroken $\scT_+$ symmetry, as illustrated in Fig.~\ref{fig:decision_tree_classes}.
We compute $O_{\alpha\balpha}^{(\scA)}$, $\scA=\scC_-,\scT_-$, in the bulk for the randomly sampled disordered spin chains of each example. The fraction of positive, negative, and zero $O_{\alpha\balpha}^{(\scC_-)}$ and of zero and nonzero $O_{\alpha\balpha}^{(\scT_-)}$ in each class are listed in Table~\ref{tab:overlap_results}. We conclude that in all of them, the criteria for $\scC_-^2$ and $\scT_-^2$ are satisfied for at least $99.9\%$ of realizations and our examples conform spectacularly to random-matrix universality, confirming the tenfold classification of many-body Lindbladians with unbroken $\scT_+$ symmetry put forward in previous sections.

\section{Discussion, conclusions, and outlook}

In this work, we put forward a symmetry classification of many-body Lindbladian superoperators and confirmed it through a study of random-matrix correlators in experimentally implementable dissipative spin chains. We found that Lindbladians without unitary symmetries and Lindbladians with symmetries in steady-state symmetry sectors belong to one of ten non-Hermitian symmetry classes. These classes are characterized by the existence of a $\scT_+^2=+1$ symmetry implemented by the \textsc{swap} operator. Going beyond sectors with steady states breaks the $\scT_+$ swap symmetry between the two copies (bra and ket) of the system and, consequently, enriches the symmetry classification.

Interestingly, we found compelling evidence that $\scC_+^2=-1$ and $\scT_+^2=-1$ symmetries cannot be implemented inside individual symmetry sectors, reducing the allowed number of classes of many-body Lindbladian from 54 to 29. This conclusion does not exclude the possibility of a $\scC_+^2=-1$ connecting different sectors. Indeed, such a symmetry can be implemented between two sectors of odd fermionic parity~\cite{lieu2022}. As a consequence, all eigenvalues are doubly degenerate, but the two eigenvalues of a given pair belong to different sectors, thus not defining a symmetry class with Kramers degeneracy.

The $\scC_+$ symmetry of a given class can be detected through the use of bulk complex spacing ratios, while $\scC_-$ and $\scT_-$ symmetries require the study of correlations on or near the axes of symmetry of the spectrum. Because of the spontaneous breaking of PT symmetry, we found eigenvalue correlations on these axes not to be useful in practice. Instead, we proposed the eigenvector overlaps between states connected by the antiunitary symmetry of interest as a useful new signature of non-Hermitian antiunitary symmetries. Importantly, they can be computed even when the explicit form of the symmetry transformation is not known. The role of these off-diagonal eigenvector overlaps as a measure of dissipative quantum chaos deserves further study. We used the sign of different overlaps as a proxy for the existence or absence of a given non-Hermitian antiunitary symmetry, but did not study in any detail their distributions. While a numerical study is the natural first step, an analytical investigation following Chalker and Mehlig~\cite{mehlig1998PRL} might be possible.

Our work complements ongoing effort to characterize PT-symmetric Lindbladian dynamics~\cite{prosen2012PRL,prosen2012PRA,vancaspel2018,huybrechts2020,huber2020,nakanishi2022,starchl2022PRL}. Note that PT symmetry is nothing but pseudo-Hermiticity of the dynamical generator. The definition of PT symmetry put forward in Ref.~\cite{huber2020}, which we would propose to call a \emph{strong PT symmetry} (or strong pseudo-Hermiticity), clearly implies the existence of a pseudo-Hermiticity transformation $\scQ_\pm$ of the Lindbladian but, by allowing for shifts of the Lindbladian spectrum, our classification goes beyond that definition and includes Lindbladians with weak pseudo-Hermiticity (\emph{weak PT symmetry}).
Remarkably, pseudo-Hermiticity has observable consequences in the transient quantum dynamics. 
More concretely, the dihedral symmetry of the spectrum implies the existence of a time-reversal-like property of certain correlation functions, despite the dynamics being dissipative.
Furthermore, if the pseudo-Hermiticity is not spontaneously broken, then there is collective decay of the eigenmodes, as all eigenvalues of the shifted Lindbladian are either purely real or purely imaginary.

Finally, our work also does not address the relation between the non-Hermitian classification of dynamical generators and the Hermitian classification of steady states. The two classifications are decoupled for quadratic open quantum systems~\cite{lieu2020}, but it is unclear, at this point, if there exists any correspondence between the Altland-Zirnbauer class~\cite{altland1997} of the steady state and the corresponding dynamical Bernard-LeClair class of the generator in the many-body case. A simple one-to-one correspondence cannot exist because Lindbladians in any of the five classes with $\scC_+^2=+1$ lead to a featureless steady state proportional to the identity [as follows from Eq.~(\ref{eq:sym_scLJ_Q+})], but there could still exist a more limited correspondence between the remaining five classes and a subset of the Altland-Zirnbauer classes. We leave a matching of symmetries on both sides (if any exists) for future work.

\textit{Note added.}---The main results of this paper were announced by one of us (L.S.) in the workshop \textit{Chaotic and
Integrable Dynamics} in Pokljuka, Slovenia on 7 July 2022. While finalizing this manuscript, we learned of a closely related work by K. Kawabata, A. Kulkarni, J. Li, T. Numasawa, and S. Ryu~\cite{kawabata2022Classes}, which appears in the same arXiv listing.

\begin{acknowledgments}%
We thank Gernot Akemann, Antonio M.\ García-García, Kohei Kawabata, Anish Kulkarni, Jiachen Li, Shinsei Ryu, Jacobus J.\ M.\ Verbaarschot, and Can Yin for illuminating discussions. This work was supported by Funda\c{c}\~ao para a Ci\^encia e a Tecnologia (FCT-Portugal) through Grants No.\ SFRH/BD/147477/2019 (LS) and UID/CTM/04540/2019 (PR). TP acknowledges ERC Advanced Grant 694544-OMNES and ARRS research program P1-0402.
This project was funded within the QuantERA II Programme that has received funding from the European Union’s Horizon 2020 research and innovation programme under Grant Agreement No 101017733.
\end{acknowledgments}

\appendix

\section{Details on the numerical simulations}
\label{app:numerics}

In this appendix, we provide additional details on the numerical sampling of spin chains and the analysis of random-matrix correlators. 

As mentioned in Sec.~\ref{sec:examples}, we always consider spin chains of $L$ sites with periodic boundary conditions and considered nearest-neighbor and next-to-nearest-neighbor interactions. More specifically, we restrict the couplings $J^{x,y,z}_{jk}$ in Eqs.~(\ref{eq:H_YXZ}) and (\ref{eq:jumpops_inchop}) to $J^{x,y,z}_{j,j+1}$, the couplings $K_{jk}$ in Eq.~(\ref{eq:H_deph_noTRS}) to $K_{j,j+1}$ and $K_{j,j+2}$, and the couplings $\gamma_{jk\ell}$ and $\eta_{jk\ell}$ in Eq.~(\ref{eq:jumpops_chiral}) to $\gamma_{j,j+1,j+2}$ and $\eta_{j,j+1,j+2}$, respectively.

To perform the statistical analysis of random-matrix correlations, we consider random 
(i.e., quenched disordered) spin chains. For a given coupling $g$, we either choose a fixed value $g=g_0$ or sample it from a box distribution in $[g_0-\d g, g_0+\d g]$, in which case we denote it as $g=g_0\pm\d g$. The values of the couplings in the different examples (in the order discussed in Sec.~\ref{sec:examples}) are as follows (we suppress the site indices, which were already discussed above):
\begin{enumerate}
    \item \textit{Dephasing, $H_\mathrm{X}$:}
    $\gamma=1.1\pm0.9$, $K=1.0$, and $g^x=0\pm2.1$.
    \item \textit{Dephasing, $H_\mathrm{XYZ}+H_\mathrm{X}$:}
    $\gamma=1.1\pm0.9$, $J^x=1.0$, $J^y=0.8$, $J^z=0.55$, and $g^x=0\pm0.7$.
    \item \textit{Dephasing, $H_\mathrm{XYZ}+H_\mathrm{X}+H_\mathrm{Y}$:}
    $\gamma=1.1\pm0.9$, $J^x=1.0$, $J^y=0.8$, $J^z=0.55$, $g^x=0\pm0.7$, and $g^y=-0.1\pm0.9$.
    \item \textit{Dephasing, $H_\mathrm{XYZ}+H_\mathrm{X}+H_\mathrm{Y}+H_\mathrm{Z}$:}
    $\gamma=1.1\pm0.9$, $J^x=1.0$, $J^y=0.8$, $J^z=0.55$, $g^x=0\pm0.7$, $g^y=-0.1\pm0.9$, and $h=0.2\pm0.3$.
    \item \textit{Spin injection and removal:}
    $a=0.8\pm0.4$, $b=0.7\pm0.5$, $J^x=1.0$, $J^y=0.8$, and $J^z=0.55$.
    \item \textit{Incoherent hopping:}
    $M^x=(0.3+0.2\i)\pm(0.2+0.5\i)$, $M^y=(0.5-0.4\i)\pm(0.4+0.1\i)$, $J^x=1.0$, $J^y=0.8$, $J^z=0.55$, and $h=3\pm2$.
    \item \textit{Incoherent hopping + dephasing:}
    $\gamma=1.1\pm0.9$, $\eta=0.4$, $K=0.8$, $h=0\pm0.7$.
\end{enumerate}

For the examples with a Liouvillian weak symmetry, we consider chains of length $L=5$, $6$, $7$, and $8$, corresponding to symmetry sectors of size $2^{2L-1}=512$, $2048$, $8192$, and $32768$, respectively. For the example with a Liouvillian strong symmetry, we also consider $L=5$, $6$, $7$, and $8$, which, in this case, correspond to sector dimensions of $2^{2L-2}=256$, $1024$, $4096$ and $16384$, respectively. For the examples without unitary symmetries, we study chains of length $L=5$, $6$, and $7$, corresponding to irreducible Liouvillians of dimension $L=2^{2L}=1024$, $4096$, and $16384$, respectively. 

The eigenvalues and eigenvectors are obtained by numerical exact diagonalization. At least $10^6$ eigenvalues were considered when computing the complex spacing ratio distribution and at least $2\times 10^6$ eigenvectors for the overlap matrix, for which we restrict ourselves to sizes $L=5$ and $L=6$. Since we are interested in the bulk correlators, we selected only the eigenvalues with both real and imaginary parts larger than $10^{-6}$ (in absolute value) and their corresponding eigenvectors.

\bibliography{main_v7.bbl}

\end{document}